\documentclass[10pt]{article}

\usepackage{graphicx,epstopdf}% Include figure files
\usepackage{epsfig}
\usepackage{latexsym}
\usepackage{amsmath}

\begin{document}
\title{\Large \bf Data science for assessing possible tax income
manipulation: the case of Italy}

\author{ \large \bf  Marcel Ausloos$^{1,2,}$\footnote {Email:
ma683@le.ac.uk,}  \footnote {Email: marcel.ausloos@ulg.ac.be}, Roy
Cerqueti$^{3,}$\footnote {Corresponding author. Email:
roy.cerqueti@unimc.it } , Tariq A. Mir$^{4,}$\footnote {Email:
taarik.mir@gmail.com}
 \\ \\$^1$   Institute of Accounting and Finance, School of Management, \\University of Leicester,
 University Road, \\ Leicester, LE1 7RH, UK
\\ $^2$ GRAPES, rue de la Belle Jardiniere, B-4031 Liege, \\Federation Wallonie-Bruxelles, Belgium
\\$^3$ University of Macerata, Department of
Economics and Law,\\  via Crescimbeni 20, I-62100, Macerata, Italy \\
\\$^4$ Nuclear Research Laboratory, Astrophysical Sciences Division, \\ Bhabha Atomic Research Center, Srinagar-190 006, \\Jammu and Kashmir, India}

\maketitle
%\end{document}
  \begin{abstract}
This paper explores a real-world fundamental theme under a data
science perspective. It specifically discusses whether fraud or
manipulation can be observed in and from municipality income tax
size distributions, through their aggregation from citizen fiscal
reports. The study case pertains to official data obtained from the
Italian Ministry of Economics and Finance over the period 2007-2011.
All Italian (20) regions are considered. The considered data science
approach concretizes in the adoption of the Benford first digit law
as quantitative tool. Marked disparities are found, - for several
regions, leading to unexpected "conclusions". The most eye browsing
regions are not the expected ones according to classical imagination
about Italy financial shadow matters.

\end{abstract}

\section{Introduction}\label{Introduction}

This paper deals with the relevant theme of identifying the
existence of anomalies in tax incomes. We specifically focus on the
case of Italian regions. The problem is faced under a data science
perspective, which is suitable for the scope of the study. Indeed,
data science represents nowadays a major area in the research
frontier for processing large sets of data (see e.g. Carbone et al.,
2016 and references therein contained).

The relevance of the study lies in the evidence that assessing the
errors in financial statements is a major task of auditors,
regulators,  or analysts not only  in financial markets, but also in
macroeconomic and public affairs, like on governmental economic
data. Reports of accurate financial statement data are crucial, even
essential, to the management of public budgets. Thus, it is
mandatory to observe whether misestimations, mistakes, biases, or
even manipulations have occurred or are occurring (Puyou, 2014). On
the other hand, academic researchers must propose ways for detecting
errors or anomalies. Many methods have been proposed and steps taken
in creating and validating techniques to assess different constructs
of errors (Cooper et al. 2013). However, despite substantial
progress,   in this  safety area, available methods present
deficiencies that limit their usefulness, - sometimes due to unclear
hypotheses underlying the method. Most likely, this will continue
for ever, since it is well known that the imagination of crooks
leads to further more sophisticated manipulation, while reaction of
"policy makers"   is impaired by legal processes.  Yet, controls are
challenged by intelligent people, lacking classical ethics.

Without suggesting opprobrium on all Italian citizens because of
supposed to be tax evasion by a few,  before debating individual
cases, it is often admitted that   Italy is one of the top countries
losing to tax evasion (after the USA and Brazil) through a GDP
ranking
($http://investorplace.com/investorpolitics/10-worst-countries-for-tax-evasion/\#.Vvkqf3AR54c$),
or through the amount of tax loss as a result of shadow economy.
Income manipulation might thus be rightfully tested on such a
(country) case, - as somewhat  in line with recently discussed
pertinent topics, from points of view related to ethics and
organized crime,  by e.g. Brosio et al. (2002), Alexeev et al.
(2004),  Calderoni (2011), or  Mir et al. (2014).

Of course, not only citizens falsify their financial data, but also
firms and even governments (Aggarwal and Lucey 2007; Michalski and
Stoltz 2013; Luippold et al. 2015).  For example, questions have
been raised about the data submitted by Greece to the Eurostat to
meet the strict deficit criteria set by the European Union (EU), -
see Rauch et al. (2011), or   about the macroeconomic data of China
(Holz 2013). Managers can engage in more or less corporate tax
avoidance than shareholders would otherwise prefer (Armstrong et al.
2015). On the other hand, firms have incentives to manipulate
earnings in order   to convince investors, e.g. to report a rounded
to a upper value number when they have profits (i.e., USD 40
million) and to report a number such as  USD 39.95 million, when
they have losses, - as discussed by Thomas (1989), having observed
such unusual patterns in reported earnings. This  rounding approach
points to a "moderate manipulation" of the data. However, its
relevance is considered not to be negligible for investors.

At the "lower level", that of citizens, it is also known that
seemingly small rounding manipulations can influence financial
statement users' perception of credit quality (Guan et al 2008). At
another level, that where the citizen is immersed in a crowd, and
expects to be protected by some shadow due to a bigger  cheater, it
is interesting to raise the question whether a collective effect can
be seen. This can be done through examining income tax contributions
at local levels. This accounting level is the core of our
investigation and report.

A review  of statistical methods of  fraud detection has been
provided by  Bolton   and Hand (2001, 2002), while  accountants'
perceptions regarding fraud detection and prevention methods have
been recently discussed  Bierstaker et al. (2006); see also  Wadhwa
and  Pal (2012) for a quick summary or  Lin et al. (2015) for a
specific discussion of a couple of techniques.

In this context, the data science approach proposed in the present
paper is based on the Benford law.

This  Benford law, originally for the first digit (BL1) distribution
of data sets,    follows a logarithmic law:
\begin{equation}
P(d)=log_{10}(1+\frac{1}{d}),\qquad d=1,2,\dots ,9,  \label{BL1}
\end{equation}%
where $P(d)$ is the probability that the  first digit is equal to
$d$ in the data set; $log_{10}$ being the logarithm in base 10.

This "law" stems from observations by Newcomb  (1881) and  later
independently by Benford  (1938)   that the distribution of the 1st
digit  is more concentrated on smaller values:  the digit "1" has
the highest frequency, "9" the lowest frequency.  In Table
\ref{tableBL1},  the frequency  of the first digit, as given by BL1,
is recalled for the reader convenience.  Thereafter, mathematics can
suggest empirical law for the 2nd, 3rd, ... digit distribution.
However, since the latter becomes quickly rather uniform, it becomes
hard (but it is done) to use  such high level digits for testing the
validity of reported data. Thus, let us concentrate our aim below to
the first digit, i.e.  on the validity (or not) of BL1 in  a
specific case, serving as a paradigm for other big data
investigations.

\begin{table}[tbp]
\begin{center}
\begin{tabular}{|l||l|l|l|l|l|l|l|l|l|}
\hline \textbf{  $d$} & 1 & 2 & 3 & 4 & 5 & 6 & 7 & 8 & 9 \\ \hline
\textbf{Freq.} & 0.301 & 0.176 & 0.125 & 0.097 & 0.079 & 0.067 &
0.058 & 0.051 & 0.046 \\ \hline
\end{tabular}%
\end{center}
\caption{Frequency ("Freq.") of the first digit ($d$) in a set of
data;  $d$ values ranging from 1 to 9,  - according to BL1,
Eq.(\ref{BL1}).}\label{tableBL1}
\end{table}

In general, Benford law has to be recommended because it contains
many advantages like not being affected  by scale invariance,  and
is admitted to be of help when there  is no supporting document to
prove  the authenticity of the "transactions" (Varian 1972).
Nowadays, this so called  "law" provides a convenient  basis for
digital analysis of sequences of numbers of similar nature. For
example,   an analysis based on Benford law has been used in a wide
variety of ways to identify instances of employee theft and tax
evasion (Guan et al 2008) and also deviation of the exchange rates
from regular paths (Carrera 2015) or of the Libor rates
(Abrantes-Metz et al., 2011).

In fact, following Thomas (1989), a "manipulation expectation" can
be obtained using the Benford (1938) law, as was later pointed out
by  Nigrini (1994, 1996, 2012).

Since Nigrini and Mittermaier (1997), it  is admitted that   BL1 can
be used to detect fraud in accounting data   reporting individual
incomes.  The presentation and demonstration of the Newcomb-Benford
law (1881-1938), as a powerful methodology in the audit field, were
further emphasized by Hill (1998), Pinkham (1961), Raimi (1985),
Durtschi et al. (2004),  among others,  and also recently in Cleary
and Thibodeau (2005), Nigrini and Miller (2012), Clippe and Ausloos
(2012), Pimbley (2014), Amiram et al. (2015), Mir (2016), Ausloos et
al. (2016). In fact, BL1 is also applied outside the financial audit
realm; e.g. see  Fu et al. (2007) for image forensics, Ausloos et
al. (2015) for birth rate anomalies, Pollach et al. (2015) for
maternal mortality rates, or elsewhere in the natural sciences
(Sambridge,  et al. 2010), and on religious activities (Mir 2012).
We notice that the literature is huge: see Alali and Romero (2013),
for approximately the last decade, Costa et al. (2013),   or Beebe
(2016),  which contains a rather exhaustive list of references.

Limiting ourselves at the public government financial
realm data tampering, let us mention, among others,   the % [42]
application of  Benford law to selected balances in the
Comprehensive Annual Financial Reports of the fifty states of the
United States (Haynes 2012; Johnson and Weggenmann 2013); or the
"political economy of numbers" international macroeconomic
statistics  (Nye and Moul  2007).  A study  on the analysis of  the
digit distribution of 134,281 contracts issued by 20 management
units in two states, in Brazil, also found significant deviations
from Benford's  law (Costa et al. 2012); see also Gava and Vitiello
(2014). Fraud detection (and prevention methods) in the Malaysian
Public Sector have been discussed in Othman et al. (2015).

At a lower scale, - ours if it has to be recalled, using Benford's
law has permitted to uncover  deficiencies in the data  reported by
local governments, like municipalities and states in several
countries,   or example,  USA or Brazil. The digit distributions of
the financial statements of  3 municipalities, Valejo City, Orange
County and Jefferson County, have been shown to have significant
departures from that expected on the basis of BL1 (Haynes 2012),

In Italy, tax collection is a fundamental source of revenues for
local governments, enabling the efficient delivery of services
(Padovani and Scorsone 2011). On the other hand, tax evasion is
known to be widespread across Italy  (Brosio et al 2002, Fiorio and
D'Amuri, 2005, Marino and Zizza 2011, Galbiati and  Zanella 2012,
Chiarini et al. 2013).

Obviously, any financial distress of municipalities, resulting from
income tax evasion,  has severe repercussions on the lives of the
taxpayers and municipal employees (Bartolini and Santolini 2012).
This is annoying for the collectivity, thus it seems important to
have some better oversight of the quality of financial statements
and accountability in view of the  demand (and use) of funds, say
returning from the Italian government. Moreover, these concerns on
data quality, on one hand, and the admittedly poor auditing
procedures  being used in Italy, on the other hand,  have
resurfaced vigorously following the bankruptcy of a number of local
government bodies during recent financial crisis   (Padovani and
Scorsone 2011), - in fact, to be fair,  more generally,  across
several industrialized countries.
 %}

Within this review of specific accounting features relevant to our
research, it might be finally  interesting to point to the reader  a
very recent and specific (by "chance") italian case, i.e. the
detection of anomalies in receivables and payables in  Italian
universities, by Ciaponi  and Mandanici (2015). This shows that
intermediary levels of  financial data scales may contain intriguing
features, whence further suggesting to raise questions on the
detection of manipulations,  through deviations from Benford law, as
here the case of   tax incomes in e.g. regions.

Thus, we have considered the aggregated values of the income tax
reports of  each of the  20 IT regions,  -  over a  recent
quinquennium: [2007-2011] for which the data is available. As
suggested by Lusk and  Halperin (2014) we calibrate our analysis
with a $\chi^2$ test.

Our point of view is politico-economic unique:  the  accounting
reliability  of the  citizen contributions to the IT GDP, -  even
though questions are numerous.    Hopefully, within this framework, one
can also (i) enlarge the knowledge of BL1 application range, (ii)
contribute to a better application of BL1 in accounting, and even
(iii) indicate that one can reach socio-economico-political
conclusions.

The paper is organized as follows: Section \ref{method} is about the
methodology.  Section \ref{8092} contains the description of the
data.  The findings are collected  in Sect. \ref{sec:results} and
discussed  in Section \ref{sec:discussion}. The last section also
allows us to  offer suggestions for further research lines. All the
Tables collecting the results at the regional level are reported in
the Appendix.

\section{Methodology}\label{method}

The  reported research here below  provides a thorough analysis
essentially  at the regional level, of the  value (= size) income
tax distribution among Italian cities,  whence their  contribution
to the country GDP. Specifically, the data of a given region is
obtained by summing the data at a municipalities level, for the
municipalities belonging to the region under examination. We refer
to the Aggregated Income Tax (AIT, hereafter) of all the citizens
living in each Italian region.

However, income manipulation by citizens, if it exists, would occur
at  the Tax Income level, whence the specific and complicated wording of the title
of this paper.

The numerical analysis is carried out on the basis of official data
obtained at the Italian Ministry of Economics and Finance (MEF), and
concerns  each year of the 2007-2011 quinquennium.   In 2011, there
were 8092  municipalities and 20 regions, with widely different
characteristics.  Notice, at once, that this   concerns  a large
number of graphs and tables. One can concatenate them, but that
surely means 20 displays of BL1 plots, - one per region, each with 5
sets of data points, one per year, as it is seen below. The first
digit location was examined using a simple home made algorithm.
Actual occurrences of each data point
was compared to expected amounts.

MS Excel was used to organize the data into a  useable form. The
original Excel data file listed municipalities according to the
alphabetical order of their names. This was pertinently useful
within Mir et al. (2016) study. The aim of the present study is to
analyze AIT data at  a mesoscale, i.e.,    regional, level. So we
first isolated and put the municipalities in their respective
regions. We used "http://www.comuni-italiani.it/nomi/index.html" to
find out the parent regions of municipalities. The Excel spreadsheet
 included  20 columns to accommodate  each year of data for each of  the  twenty regions and  six columns for the  five  years of
analysis plus their average over the quinquennium. The first digits were extracted by using the LEFT($text$, $num_{-}chars$) function of Excel by entering the column number of interest at "$text$" and "1" (for first digit) at "$num_{-}chars$". The frequency of each digit, 1 to 9, as first digit is obtained by making use of COUNTIF($range$, $criteria$) function by specifying the "$range$" of cells and the digit of interest at "$criteria$".   The frequency of each digit, 1 to 9, as the first  significant digit is thus determined.

Several tests can be made in order to assess the validity of BL1.
The most classical $\chi^2$ test is used as for other BL1
applications (Durtschi et al. 2004).
Nevertheless, within this
constraint,
the visualization of the data and the reported test tables allow  us to pin point regularities or anomalies:
''acceptable conformity'', suggests that the balance is likely not
biased and should be accepted without further analysis;
''nonconformity''  does not guarantee that problems exist in the
underlying accounts comprising the AIT or that fraud has occurred,
but results of the BL1 analysis should be used as an indicator that
further investigation is needed.  Indeed if the data does not
conform to BL1, this signals that the aggregated data may not be
true representations; the numbers may be influenced by operations,
biased due to (our, but we did much cross checking) error, or they
may have been manipulated to deceive some financial statement user.

From the ergodicity point of view (Palmer 1982,  Lucas and Sargent
1981; Davidson 1987, 1996, 2009; Samuelson 1969; Tsallis et al.
2003), it is of interest to assess the data  through a time average;
this has been taken over the relevant 5 years for each region. A
$\chi^2$  test has been made also on such averages.

\section{Data}\label{8092}

The economic data analyzed  here below has been obtained by (and
from) the Research Center of the Italian MEF.   Contributions
have been  disaggregated at the  municipal level (in IT a
\textit{municipality} or \textit{city} is denoted as
\textit{comune}, - plural $comuni$) to the Italian GDP, for  five
recent years: 2007-2011.

Let it be recalled  that Italy  is composed of 20 regions and more
than  8000 municipalities: the latter number has varied over time,
even during the examined quinquennium:  from 8101 down to 8092,
between 2007 and 2011: several (10) cities have  merged into  (3)
new entities,  while (2) others were phagocytized. Moreover,   7
municipalities have changed  from the Marche a region to another one
(Emilia-Romagna), in 2008. The variation in the  number of cities in
each concerned region has been taken into account: we have made   a
virtual merging of cities   (see also
$http://www.comuni-italiani.it/regioni.html$), in order to compare
AIT  data for  "stable size" regions. Since the number of regions
has been constantly equal to 20,  the regional level   seems to be
the most  interesting one for any data measure and discussion. The
regions are listed by order of importance, i.e. through their number
of cities, in Table \ref{TableNcityperregion}. Observe that the
to-be-expected $\chi^2$ calculated for the 2011 region municipality
number content is given for future reference, and avoiding extra
columns or lines in other Tables.

In short, the AIT of the resulting cities,  whence that of the
regions,   was linearly adapted, as if these cities and regional
content were preexisting before the merging or phagocytosis. The
(rounded) AIT  in successive years and  the corresponding averaged
AIT of each region over the quinquennium  are given in Table \ref
{tableAITperreg}. The AIT and $<$AIT$>$,   for IT  are in (e+11)
EUR,  but for regions in (e+10) EUR.  As a complementary
information,  but irrelevant for the BL1 test,   the number of
inhabitants ($N_{inhab}$) in each region, according to the 2011
Census, is also reported  from
$http://dati-censimentopopolazione.istat.it/Index.aspx?lang=en$.

\begin{table} \begin{center}
\begin{tabular}[t]{|c|c|c|c|c|c|}
  \hline
  &\multicolumn{3}{|c|}{$year$}  & (*)& \\ \hline
     &2007&2009&2011 & 2011 &  \\ \hline
  $ region$&\multicolumn{3}{|c|}{$N _{c,r}$} &$N_{inhab}$&$<\chi^2 >$ \\ \hline
Lombardia   &   1546    &   1546    &   $\downarrow$    1544    &   9 704 151   &   6.3694   \\ \hline
Piemonte    &   1206    &       &           &   4 363 916   &   5.2830  \\  \hline
Veneto  &   581 &       &           &    4 857 210  &   7.7896  \\  \hline
Campania    &   551 &       &           &   5 766 810   &   17.224  \\  \hline
Calabria    &   409 &       &           &   1 959 050   &   4.4664  \\  \hline
Sicilia &   390 &       &           &   5 002 904   &   6.4494  \\  \hline
Lazio   &   378 &       &           &    5 502 886  &   5.0288  \\  \hline
Sardegna    &   377 &       &           &   1 639 362   &   24.587  \\  \hline
Emilia-Romagna  &   341 &   341 &   $\uparrow$  348 &   4 342 13    &   8.1620  \\  \hline
Trentino- Adige Alto    &   339 &$\downarrow$   333 &       333 &   1 029 475   &   5.3782  \\  \hline
Abruzzo &   305 &       &           &   1 307 30    &   4.7372  \\  \hline
Toscana &   287 &       &           &   3 672 202   &   9.5140  \\  \hline
Puglia  &   258 &       &           &   4 052 566   &   5.8868  \\  \hline
Marche  &   246 &   246 &   $\downarrow$    239 &   1 541 319   &   8.6922  \\  \hline
Liguria &   235 &       &           &    1  570 694 &   16.895  \\  \hline
Friuli- Giulia Venezia  &   219 &   218 &       218 &   1 218 985   &   7.8444  \\  \hline
Molise  &   136 &       &           &    313 660    &   12.657  \\  \hline
Basilicata  &   131 &       &           &   578 036 &   7.5054  \\  \hline
Umbria  &   92  &       &           &   884 268 &   11.317  \\  \hline
ValledAosta &   74  &       &           &    126 806    &   7.6682  \\  \hline
Total   &   8101    &$\downarrow$   8094    &   $\downarrow$    8092    &   59 433 744  &       \\  \hline
\end{tabular}
\caption{Number $N _{c,r}$ of  cities in 2011, and in previous
years, in  the (20) IT regions; the region ranking follows the
decreasing city number.   For  a more complete information on IT
cities,  the number of inhabitants ($N_{inhab}$) in each region,
according to the 2011 Census, is also reported (*)
$http://dati-censimentopopolazione.istat.it/Index.aspx?lang=en$.
 The empirical  $ \chi^2 $ of  the averaged BL1 observations for
each region can be read in the last column: it is included in this
Table in order to alleviate other Tables.  Also recall  that the
critical value of the $\chi^2$ test at a confidence level 0.05 is
$\chi^2_{0.05} = 15.51$ for a number of  "degrees of freedom"  $\partial
=8$, which is that of BL1.
 }.
 \label{TableNcityperregion}
\end{center} \end{table}

 \begin{table}\label{summarytable}
%  \begin{center}
  \begin{centering}
  \begin{tabular}{|c|c|c|c|c|c|c|c|c|c|c|c| }
\hline
$i$&$ N_{c,r}$   &       in    &\multicolumn{4}{|c|}{AIT}   &&  $<$AIT$>$    \\
= &     &    REGION & 2007   &      2008 &   2009    &   2010 & 2011  &  5yr   \\ \hline\hline
1&  305 &   ABRUZZO &   1.2871  &   1.2850  &   1.3053  &   1.3288  &   1.3613  &   1.3135     \\
2&  131 &   BASILICATA  &   0.4580  &   0.4735  &   0.4820      &   0.4834  &       0.4911      &   0.4776     \\%587 717 \\
3&  409 &   CALABRIA    &   1.3404  &   1.3961  &   1.4411  &   1.4496  &   1.4516  &   1.4158   \\
4&  551 &   CAMPANIA    &   4.1890  &   4.2908  &   4.3589  &   4.3833  &   4.3863  &   4.3217  \\
5&  348 &   EM. ROMAGNA &   6.2211  &   6.3448  &   6.2945  &   6.3665  &   6.3970  &   6.3248  \\
6&  218 &   FRIULI V.G. &   1.6876  &   1.7121  &   1.7163  &   1.7214  &   1.7323  &   1.7139   \\
7&  378 &   LAZIO   &   7.1759  &   7.3436  &   7.4487  &   7.5532  &   7.6163  &   7.4275   \\
8&  235 &   LIGURIA &   2.2020  &   2.2402  &   2.2829  &   2.2958  &   2.3003  &   2.2642   \\ %\\
9&  1544    &   LOMBARDIA   &   14.457  &   14.737  &   14.561  &   14.771  &   15.008  &   14.707   \\
10& 239 &   MARCHE  &   1.7710  &   1.8045  &   1.7977  &   1.8232  &   1.8567  &   1.8106   \\
11& 136 &   MOLISE  &   0.2736      &   0.2823  &   0.2825      &   0.2827      &   0.2865      &   0.2815  \\
12& 1206    &   PIEMONTE    &   5.9479  &   6.0326  &   5.9797  &   6.0515  &   6.1201  &   6.0264  \\%4 393 838\\
13& 258 &   PUGLIA  &   3.1445  &   3.2563  &   3.3082  &   3.3557  &   3.3947  &   3.2919   \\
14& 377 &   SARDEGNA    &   1.4896  &   1.5510  &   1.5789  &   1.5890  &   1.5977  &   1.5612  \\
15& 390 &   SICILIA &   3.6977  &   3.8324  &   3.9066  &   3.9256  &   3.9451  &   3.8615  \\
16& 287 &   TOSCANA &   4.7404  &   4.8175  &   4.8417  &   4.8943  &   4.9499  &   4.8487   \\
17& 333 &   TRENTINO A.A.   &   1.3967  &   1.4379  &   1.4808  &   1.5148  &   1.5360  &   1.4733    \\
18& 92  &   UMBRIA  &   1.0167  &   1.0432  &   1.0539  &   1.0624  &   1.0702  &   1.0493   \\
19& 74  &   V. D'AOSTA  &   0.1795  &   0.1849  &   0.1889  &   0.1911  &   0.1923  &   0.1873   \\ %126 620\\
20& 581 &   VENETO  &   6.2346  &   6.3244  &   6.2912  &   6.3808  &   6.4845  &   6.3431   \\\hline
IT & 8092    &   & 6.8910 & 7.0390 &    7.0601 &    7.1424 &    7.2178 &    7.0701    \\\hline
 %T**** & 8092   &???  & 6.8910 & 7.0390 &  7.0601 &    7.1424 &    7.2178 &    7.0701 & 60 545 909\\\hline
 \end{tabular} %  \end{center}
 \caption{$ N_{c,r}$,  the number of  cities in each region;
(rounded) AIT  in successive years and  average AIT of a region over
the quinquennium.  The AIT and $<$AIT$>$,   for IT is in (e+11) EUR,
but for regions in (e+10) EUR. } \label{tableAITperreg}
\end{centering}
\end{table}

\begin{table} \begin{center}
\begin{tabular}[t]{ccccccccc}
  \hline
   $ $   &2007 &2008 &2009&2010&   2011 & $<$AIT$>$ &\\
\hline
Min. (/e+09)     &  1.7950   &  1.8495   &  1.8891   &  1.9110   &  1.9230   &  1.8735  &\\
Maxi (/e+10)     &  14.457   &  14.737   &  14.561   &  14.771   &  15.008   &  14.707  &\\
Sum  (/e+10)     &  69.910   &  70.390   &  70.601   &  71.424   &  72.178   &  70.701  &\\
%Points  &  20   &  20   &  20   &  20   &  20   &  20  &\\
Mean (/e+10)     &  3.4455   &  3.5195   &  3.5300   &  3.5712   &  3.6089   &  3.5350  &\\
Median (/e+10)   &  1.9865   &  2.0223   &  2.0403   &  2.0595   &  2.0785   &  2.0374  &\\
RMS  (/e+10) &  4.7793   &  4.8753   &  4.8601   &  4.9227   &  4.9831   &  4.8840  &\\
Std Dev  (/e+10)     &  3.3982   &  3.4613   &  3.4274   &  3.4761   &  3.5254   &  3.4576  &\\
%Variance(/e+21)     &  1.1548   &  1.1980   &  1.1747   &  1.2084   &  1.2429   &  1.1955  &\\
Std Error (/e+09)    &  7.5986   &  7.7397   &  7.6639   &  7.7729   &  7.8831   &  7.7314  &\\
Skewness     &  1.7795   &  1.7784   &  1.7406   &  1.7478   &  1.7672   &  1.7627  &\\
Kurtosis     &  3.4321   &  3.4359   &  3.2850   &  3.3097   &  3.3902   &  3.3708  &\\\hline
% $\mu/\sigma$  &0.1309   &0.1318&0.1321 &0.1319&  0.1329&  \\
%$3(\mu-m)/\sigma$  &0.2873&0.2884&0.2883 &0.2878& 0.2889&   \\
%\hline
\end{tabular}
   \caption{  Summary of  (rounded) statistical
characteristics  for AIT (in  Euros) of the   IT   regions
($N_r=20$) in 2007-2011.  }\label{Tablestatmircond}
\end{center} \end{table}

 \section{Results} \label{sec:results}
Beside  the  (rounded) AIT  in successive years and  average AIT of
a region over the quinquennium  AIT and $<$AIT$>$,  given for  the
regions in (e+10) EUR)  units, and for IT  in  (e+11) EUR,   given
in Table \ref{tableAITperreg}, the statistical characteristics  of
the AIT  regional distribution    for  2007-2011 is  reported in
Table \ref{Tablestatmircond},    together  with the corresponding
time average. The skewness and kurtosis are obviously both positive,
and the mean greater than the median (by a factor $\sim 1.75$).
Relevantly, it can be observed that the minimum and maximum AIT,
both have 1 as first digit.

\subsection {BL1 displays}\label{figs}

Each BL1 data set for each (20) region is displayed on Figs. 1-20;
different symbols are used in order to distinguish the 5 years so
examined. On each figure, the frequency of the $d$ digit is given,
together with the theoretical BL1. Moreover, the sample standard
error bar, i.e. depending on the number of cities, i.e. $\sigma
\simeq (1/\sqrt{ (N _{c,r}-1)}) $ allows to draw the estimated range
of the confidence interval defined as  [BL1-, BL1+], in obvious
notations. The display order has been chosen according to the region
ranking given  in Table \ref{TableNcityperregion}.

Denote the first digit as $d$, so that $d=1, \dots, 9$. Visual observations (list in descending order of $N_{c,r}$) point to:
\begin{itemize}
\item
Lombardia: rather smooth distribution,
not too far from confidence interval  (CI);
$d=6, 7$ somewhat away from CI; above BL1+.%& 1546     &1546  &   $\downarrow$    1544&& 6.3694
  \item
Piemonte: rather smooth distribution, not too far from CI;
$d=5, 6$ somewhat away from CI, specifically above BL1+.  % &1206&& & &5.2830
 \item
Veneto: rather smooth distribution, not too far from CI, even if
$d=4,5$ are somewhat away from CI, on different sides of BL1.
$d=9$ is much dispersed. % &581&& & & 7.7896
\item
Campania: very scattered, except $d=1, 2$;
other digits are far from CI. %&   551&& & & 17.224
 \item
Calabria: rather smooth distribution, not too far from  CI;
$d=6, 7$ somewhat away from CI, above BL1+; the digit $d=2$ is below BL1-. %&   409&& & & 4.4664
\item
Sicilia: $d= 1, 4$ below BL1-; $d=7$ much dispersed. %&    390&& & &  6.4494
\item
Lazio: rather smooth distribution, not too far from CI, even if
$d=4, 5, 6$ are somewhat away from CI (below BL1-).%   &378&& & &5.0288
 \item
Sardegna: very very scattered, except for $d=1,2$. %    &377&& & &  24.587
\item
Emilia-Romagna: rather smooth distribution, not too far from CI.
$d=3, 4, 5 $ are somewhat away from CI  and below BL1-. The digit
$d=8$ is quite above BL1+. %&  341&341    & $\uparrow 348$ & & 8.1620
 \item
Trentino-Alto Adige: rather smooth distribution, not too far from CI,
but  $d=  4  $ somewhat away from CI (above BL1+). $d=7$ is quite above BL1+, while $d=6$ somewhat below BL1-.
  %&  339   &$\downarrow$   333&333&&  5.3782
\item
Abruzzo:  rather smooth distribution,  but scattered away from CI;
on different sides of BL1. %&    305&& & &4.7372
 \item
Toscana: much dispersed; $d= 3, 4, 6, 7$ below BL1-; $d=8, 9$
quite above BL1+. % &287&& & &9.514
 \item
Puglia: much dispersed; $d= 3,   6, 7$ much below BL1-;
the digits $d=2, 9 $  are  quite above BL1+. %& 258&& & & 5.8868
\item
Marche:  very much dispersed; the digits $d= 3,   6 $ are much below
BL1-, while $d=4, 5, 8, 9 $  are  quite above BL1+.  %  &246&246    &$\downarrow$   239&  & 8.6922
\item
Liguria:  very much dispersed. The digits $d= 3,   6, 8 $ are
much below BL1-, while $d=2, 4, 6, 9 $ are   quite above BL1+.   %&    235&& & &16.895
 \item
Friuli-Venezia Giulia : rather smooth distribution, but rather far from CI;
$d=4, 5, 7$ quite away from CI ,  below BL1-; $d=2,3,8$ rather away above
BL1+. %& 219  &$\downarrow$   218 &218& &  7.8444
\item
Molise: very much scattered from BL1. The digits $d= 1, 2$ are much
much below BL1-, while $d=3, 6$ are much much above BL1+.
Wide variation for other digits. %  &136&& & &12.657
\item
Basilicata:   much scattered from BL1. The digits $d= 4,8$ are much below
BL1-, while $d=5, 6$ are  much above BL1+. A wide variation for all digits is observed.  %&131&& & & 7.5054
\item
Umbria:   much scattered from BL1. The digits $d= 4,6, 7, 8, 9$ are
much below BL1-, while $d=1, 2, 5$ are  much above BL1+. Also in this case,
wide variation for all digits  %& 92&& & &11.317
\item
Valle d'Aosta:   much scattered from BL1. $d= 4, 8 $ are much below BL1-, and
$d=  2, 5, 7$ are  much above BL1+. Once again, there is wide variation
for all digits  %&  74&& &  &7.6682
\end{itemize}

\subsection {  $\chi^2$ conformity test}\label{chi2}

It has been emphasized that there are five  $\chi^2$ values to be
calculated for each region, - one for each year.  These 1one hundred  $\chi^2$
values are found in Tables
\ref{TableNcityperregion1}-\ref{TableNcityperregion3}. Moreover,
the resulting $\chi^2$  for the time average values of the AIT is
also given,- but  recalled that it is given in Table
\ref{TableNcityperregion} for simplifying the reading of the  Tables
\ref{TableNcityperregion1}-\ref{TableNcityperregion3}.

However,  before examining each region BL1, it  seems  fair to
examine  whether the distribution of the $\chi^2$ itself  is
anomalous, - in order to pin point (if it occurs) some statistical
anomaly arising from computations. Such a distribution of the 100
(20 regions, 5 years)  $\chi^2$ is shown in Fig. 21.

The distribution is markedly  positively skewed, but rather
irregular, with a set of outliers. Recall that the critical value of
the $\chi^2$ at a significance level $0.05$ is $\chi^2_{0.05} =
15.51$ when the number of degrees of freedom is $\partial =8$ - that
of BL1. This rather highly peaked distribution with a few outliers
(in particular Sardegna  and Liguria, - noticed in different fiscal
years) is another hint toward pursuing a BL1 analysis at the
regional basis.

For emphasizing the regional aspects, and connecting with Table
\ref{TableNcityperregion} last column, the distribution of  the
$mean$ (average over the quinquennium) $<\chi^2>$ for each region,
is shown in  Fig. 22. Three regions are markedly outliers in the
upper range: Sardegna, Campania and  Liguria,  for  approximately
$<\chi^2> \ge 15$, pointing to much (questionable) non-conformity
with BL1.  On the contrary, 2 regions have a low $<\chi^2>$ over the
quinquennium,  Abruzzo and Calabria, indicating  very fine agreement
with BL1 ($<\chi^2> \le 4$).

At this stage, it  seems important to point to a quite substantial time-invariance of the   $ \chi^2 $   values, according to the displayed data/.

Thereafter,  it seems natural to discuss each region BL1   values,
to notice conformity or not,  starting from the most valid and
ending with the most anomalous. In so doing, one can distinguish
values according to the standard risk value for rejecting the null
hypothesis, i.e. a uniform distribution,   at $\chi^2_{0.05} =
15.507$ for a number of degrees of freedom $\partial =8$;  for
completeness, let us observe that (for   $\partial =8$) the
critical value of the $\chi^2$ at a significance level $0.10$ is
$\chi^2_{0.10} = 13.362$.

 \subsubsection{$<\chi^2 >  \le 13.362$, and yearly (ir)regularity}
It is found from Table \ref{TableNcityperregion} that the BL1
$<\chi^2>$  has a small  value in   17  regions;    from the lowest
to the highest:
 Calabria (4.4664),
 Abruzzo (4.7372),
 Lazio (5.0288),
  Piemonte (5.2830)
  Trentino-Alto Adige (5.3782),
 Puglia (5.8868)
  Lombardia (6.3694)
 Sicilia (6.4494)
 Basilicata (7.5054),
  Valle d'Aosta (7.6682),
  Veneto (7.7896),
  Friuli-Venezia Giulia (7.8444),
  Emilia-Romagna (8.1620),
 Marche (8.6922),
 Toscana (9.5140), Umbria (11.317) and Molise (12.657).

Among these, Tables
\ref{TableNcityperregion1}-\ref{TableNcityperregion3} show that much
regularity is observed in the respective yearly values, with some
exceptions.

The values for Veneto (11.33) in 2007, Emilia-Romagna (10.60) in
2010, and Valle d'Aosta (11.31) in 2010 are such that the respective
$\chi^2$ fall outside the 5\% sampling error bar. The largest
$\chi^2$ value  for  Molise occurs  in  2008,   $\sim$ 20.99,
although  a surprisingly quite small  $\chi^2$ value $\sim$ 9.74
occurs in 2007. The $\chi^2$ values  for  Umbria are high, but
without any  severe hint of an anomaly; the distribution of $\chi^2$
values is quite narrow indeed  for Umbria,  implying some
systematicity.

          \subsubsection{$<\chi^2>   \ge  15.507$, and yearly (ir)regularity}
In contrast, Liguria (16.895),  Campania (17.224), and   Sardegna
(24.587)  are  the 3 regions with an indication of much lack of
conformity  with respect to  BL1.

The  two largest  $\chi^2$ values for  Sardegna  occur  in 2007 and
2011:  $\sim$ 56.00 and 21.36, respectively. The largest  $\chi^2$
values for  Campania  occur  in 2007 and 2008: $\sim$ 21.65 and
23.02, respectively. The largest  $\chi^2$ value  for  Liguria
occurs  in  2008: $\sim$ 27.17; surprisingly, a  quite small
$\chi^2$ value $\sim$ 9.70 occurs in 2011.

\section{Discussion}\label{sec:discussion}

This section fixes and discusses the results of the investigation.

In general, the concordance between the AIT of Italian regions and
the theoretical statement of BL1 is rather questionable. There are
discrepancies at a regional level, an this is in line with the
heterogeneous nature of Italian regions under a socio-economic point
of view. In particular, one can note a very good matching between
geographic and economic features of the regions, and cluster them
among $N$ (North), $C$ (Center), $S$ (South, plus Sicilia and
Sardegna).

$N$ is the part of Italy constituted by 8 regions: Emilia-Romagna,
Friuli-Venezia Giulia, Liguria, Lombardia, Piemonte, Trentino-Alto
Adige, Valle d'Aosta and Veneto;

$C$ contains 5 regions: Abruzzo, Lazio, Marche, Toscana and Umbria;

$S$ is the remaining  7 regions:   Basilicata, Calabria, Campania,
Molise, Puglia, Sardegna, Sicilia.

Deviations from BL1 are usually read as ''data manipulation''.
However, the law   is surely a subject of controversy in accounting.
It is not clear even now neither why it should be valid at all,
under whatever "socio-economic conditions", nor whether its
theoretical derivation, under various hypotheses, informs us on its
origins and its range of applications. Even Newcomb and Benford were
dubious of the realm of validity. Therefore, a deep exploration of
the regional reality behind the AIT data, of how they have been
collected and of the shadow economy at a regional level are required
to provide a rigorous interpretation of the results. This is
well-beyond the scopes of this paper. We can only give some
suggestions and discussions, to be likely taken as arguments  for future studies.

Among the 3  regions with very anomalous BL1 $ \chi^2 $, two  belong
to $S$: Sardegna and Campania, while the other comes from $N$: Liguria.

Sardegna is characterized by a noticeable fragmentation at a city
level in several municipalities with very small number of
inhabitants. This region does not have a highly developed industrial
structure, and a wide part of the regional economy is still based on
agriculture and livestock. In the small communities the economy is
somewhat closed, and business exchanges are often based on
commodities. In such a situation, one can guess that  the existence of a
discrepancy between the official data and the income tax  should
come from the real regional economy.

Sadly, Campania has the relevant problem of a massive influence of
the organized crime on the economic system. Hence, deviations from
BL1 can be viewed "as  expected''.

The economic system of Liguria seems to be affected by the pervasion
of shadow economy. Confartigianato (the Italian association of
artisans and small businesses) states that about 73\% of the
artisans is in competition with illegal and shadow economies (see
$http://www.confartigianatoliguria.it/node/4153$). This evidence
represents a good hint for explaining why Liguria exhibits this
discrepancy with respect to BL1.

Unexpectedly, we admit so, the remaining regions are in accordance
with the BL1, with some disparities over the quinquennium as
highlighted in the previous Section. From both an economic and a
social point of view, some regions are quite similar to Campania,
with a remarkable presence of organized crime (think at Calabria,
Puglia and Sicilia, but also to the North with Veneto and
Lombardia). Basilicata and Molise are similar to Sardegna for what
concerns the absence of a well-established industrial structure. It
is also worth mentioning that Basilicata is the main producer of
fossil fuels in Italy (see the report in
$http://www.siteb.it/new\%20siteb/documenti/RASSEGNA/6711_7.pdf$).
Moreover, shadow economy is generally widespread in the entire
country.

To conclude: we demonstrate  some hints for further exploring  cases of
violations of BL1, whence likely
possible tax income manipulation through accounting city
Aggregated Income Tax reports  throughout all Italian Regions.
We admit that not every finding can be explained only based on BL1: we do not understand why some regions do not
 have  BL1 violators; we avoid  to propose speculative statements which might be called "resulting from imagination". Yet, this further supports the point  that a deeper analysis should be carried out to investigate the nature of the
fiscal data and how they are usually collected and approved (Pentland  and Carlile 1996).

\section{Conclusions}\label{conclusions}

Today Benford's law is routinely used by forensic analysts to detect
error, incompleteness and dubious  manipulation of  financial data.
The basic premise of the test is that the  first digits in real data, in
general, have a tendency to approach the Benford distribution
whereas people intending to play with the numbers, when unaware of
the law, try to place the digits uniformly. Thus any departure from
the law raises some suspicion. We have assessed  the tax income
possible manipulation  of citizens in Italy through  accounting city
aggregated income tax  reports  from all  Italian regions, with data
obtained  from  the Research Center of the Italian MEF.

The validity of the reported data  does not seem to have attracted
official accountants. For example, something like {\it Economia e
Finanza locale Rapporto 2010} or {\it RAPPORTO ANNUALE 2012 La
situazione del Paese} fall  short of discussing the data validity.

This paper provides an examination of a fiscal data set stemming
from the Italian citizens, on a regional level. Specifically, it
focuses on the assessment of potential manipulation of tax income
through the adoption of the Benford law for the first digit over the
quinquennium 2007-2011.

The BL1  presents  significant advantages over alternative measures
of accounting quality currently used in  practice. For example, it
does not require time-series, cross-sectional, or forward-looking
information, nor details on "transactions".

Throughout the paper we refer to municipalities, though in practice
we are investigating the incomes of  citizens, but we  avoid any
individual information on whether individuals correctly report their
income. This is an important distinction to the extent that  this is
precisely the population set of interest. Though we find significant
variations in municipality tax incomes by regions, much of the
variation is actually attributable to differences in the
 characteristics of  regions

Another purpose of this paper  has been to provide a proof of  the
BL1 concept for using the fiscal data at a regional level in order
to provide some information on manipulation. It is shown that it is
possible to document the variation in income taxes across regions,
to order them, and to observe a distribution of anomalies, also in
time. The  sampling data at this aggregated level is   large enough
to look in details at pertinent numbers, - without making strong
econometric modeling assumptions.   The restriction on the demand
of a large database (at the country level) expected to provide the
scale needed for the data to be sufficiently   granular can be
relaxed. The data analysis  points to different regional realities,
sometimes quite unexpectedly.

%{\it FROM Wagg \&   : Governments and government related not-for-profit entities are accounted for differently than for-profit entities. State and local governments follow reporting standards set forth by the Governmental Accounting Standards Board (GASB), while for-profit companies and non-governmental not-for-prrofits follow standards set forth by the Financial Accounting Standards Board (FASB). Governments are also subject to Government Accountability Office (GAO) audit standards and to Office of Management and Budget (OMB) directives for pass through expenditures. It is recommended, but not required, that state governments issue a CAFR, or comprehensive annual financial report (American Institute of Certified Public Accountants, 2000).}
Durtschi et.al (2004) have pointed out that when interpreting
results of Benford's  law, one should be aware of  a few risks:
%(1) False Positives-Benford shows nonconformance when the data conforms (Type I or efficiency error)
%(2). False Negatives-Benford shows conformance when the data does not conform (Type II or effectiveness error). According to Durtschi, et.al (2004),
However, Benford law is most effective   for large data set,  when
data represents more than one distribution,  when the mean is
greater than the median,  and the skewness is positive. This seems
to be the case in our investigation.

Our findings  demonstrate also that Benford's law  seems effective
in detecting data bias in not too large data sets. There is (alas)
no doubt that manipulation of income reports exist in various
regions and municipalities. Either a few accountants are so well
aware that BL1 is a test and can avoid the non-conformity surprise,
at the individual level, but cannot do so at the next (city
aggregation) level.

Thus, to  our knowledge, this paper is the first to document whether
income tax aggregated data conforms to the (first) Benford's law,
i.e. how  without examining (individual) citizens financial reports
are likely to exhibit divergences. In so doing,  a  view of fraud or
manipulation is put on a more collective level.

\subsection*{Acknowledgements}
This paper is part of scientific activities in COST Action IS1104, "The EU in the new complex geography of economic systems: models,
tools and policy evaluation".

\section*{Appendix: figures and big tables}
\begin{figure}[t]
\hspace*{-45pt}
\vspace*{-65pt}
  \includegraphics[width=.77\linewidth, angle=270]{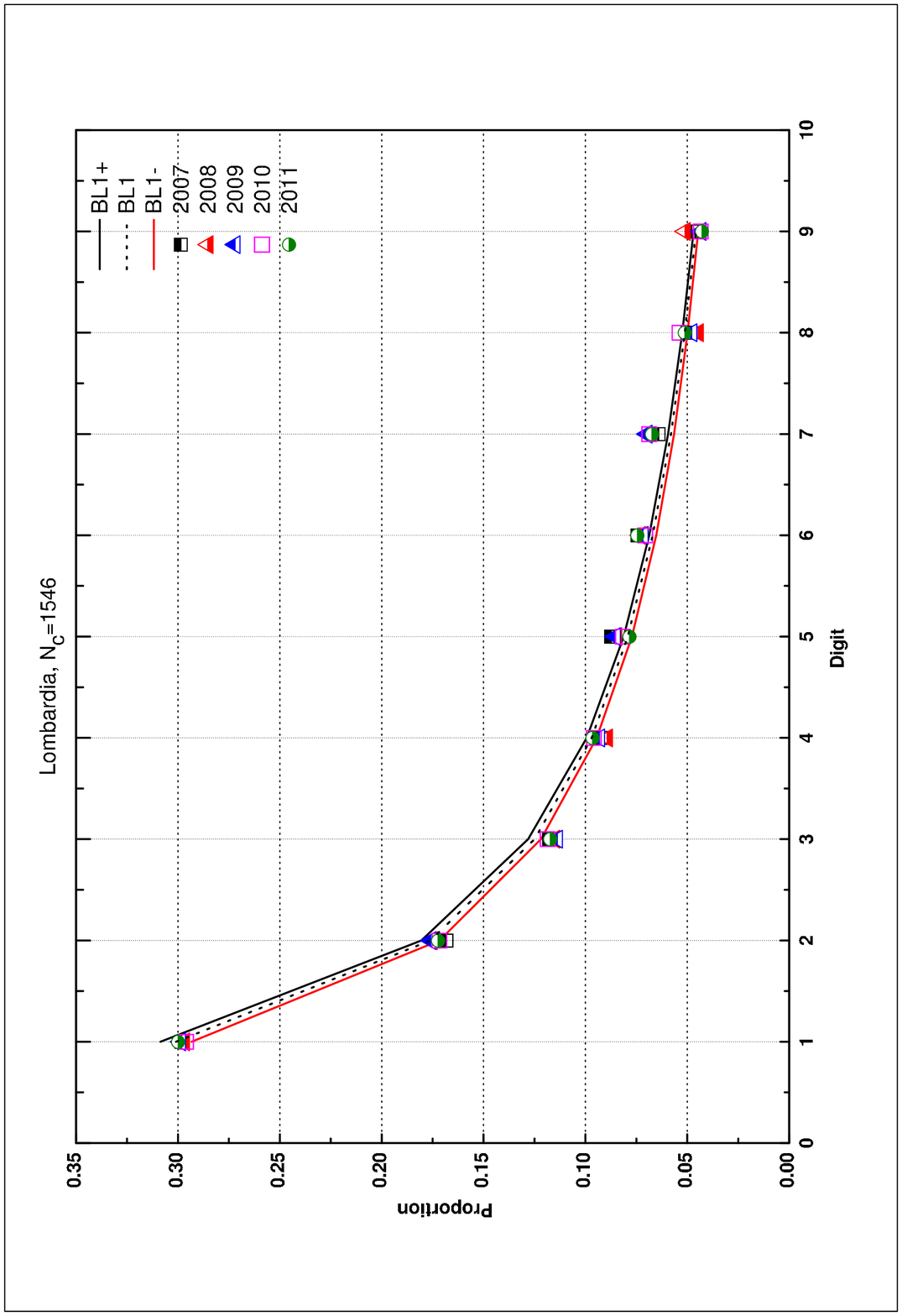}
\vspace*{50pt}  
  \caption{Lombardia.}
  \label{fig:LOMBARDIA}
\hspace*{-45pt}
  \includegraphics[width=.77\linewidth, angle=270]{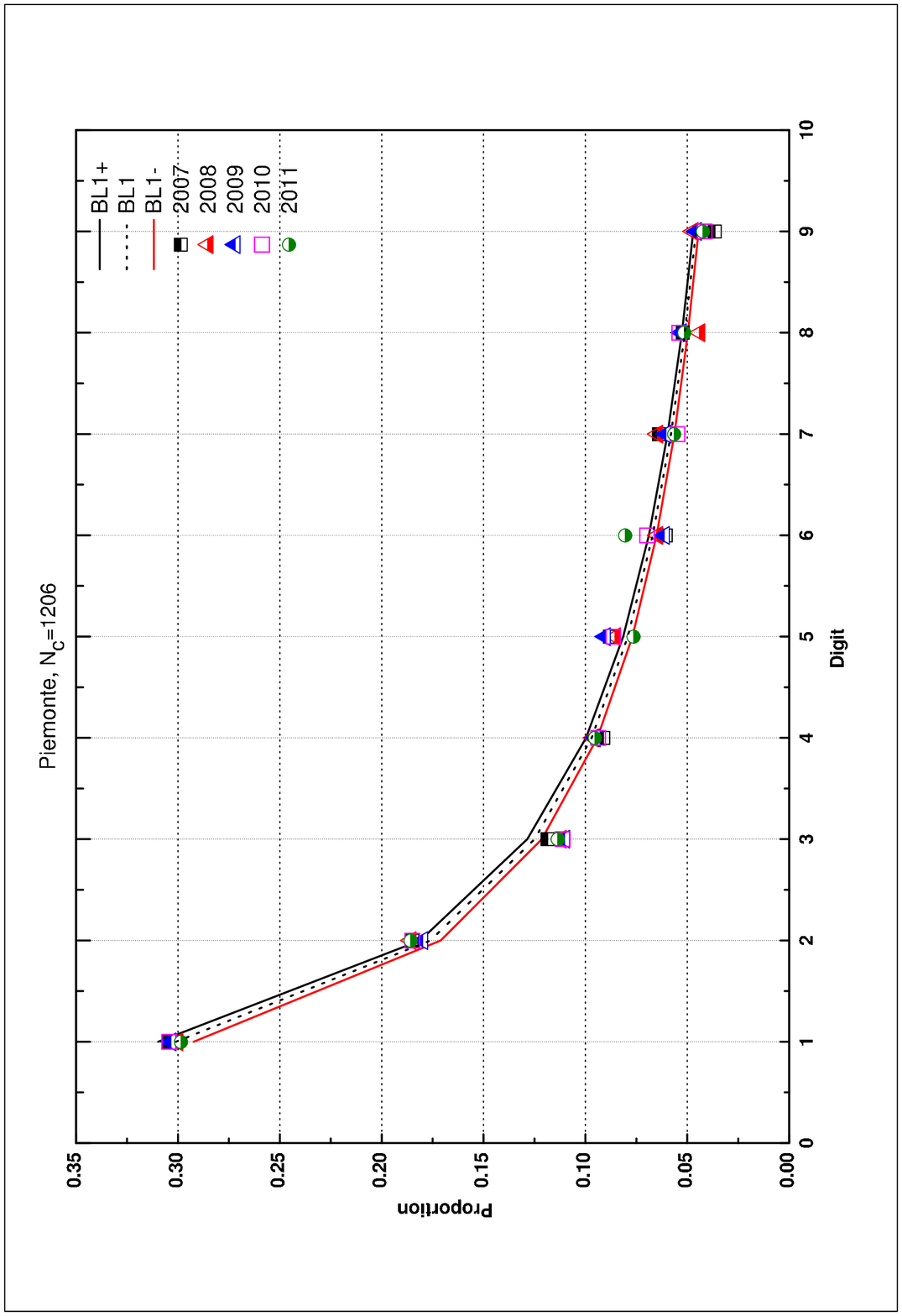}
\vspace*{-10pt}
  \caption{Piemonte.}
  \label{fig:PIEMONTE}
  %\end{center}
\end{figure}

\begin{figure}[t]
\hspace*{-45pt}
\vspace*{-65pt}
  \includegraphics[width=.77\linewidth, angle=270]{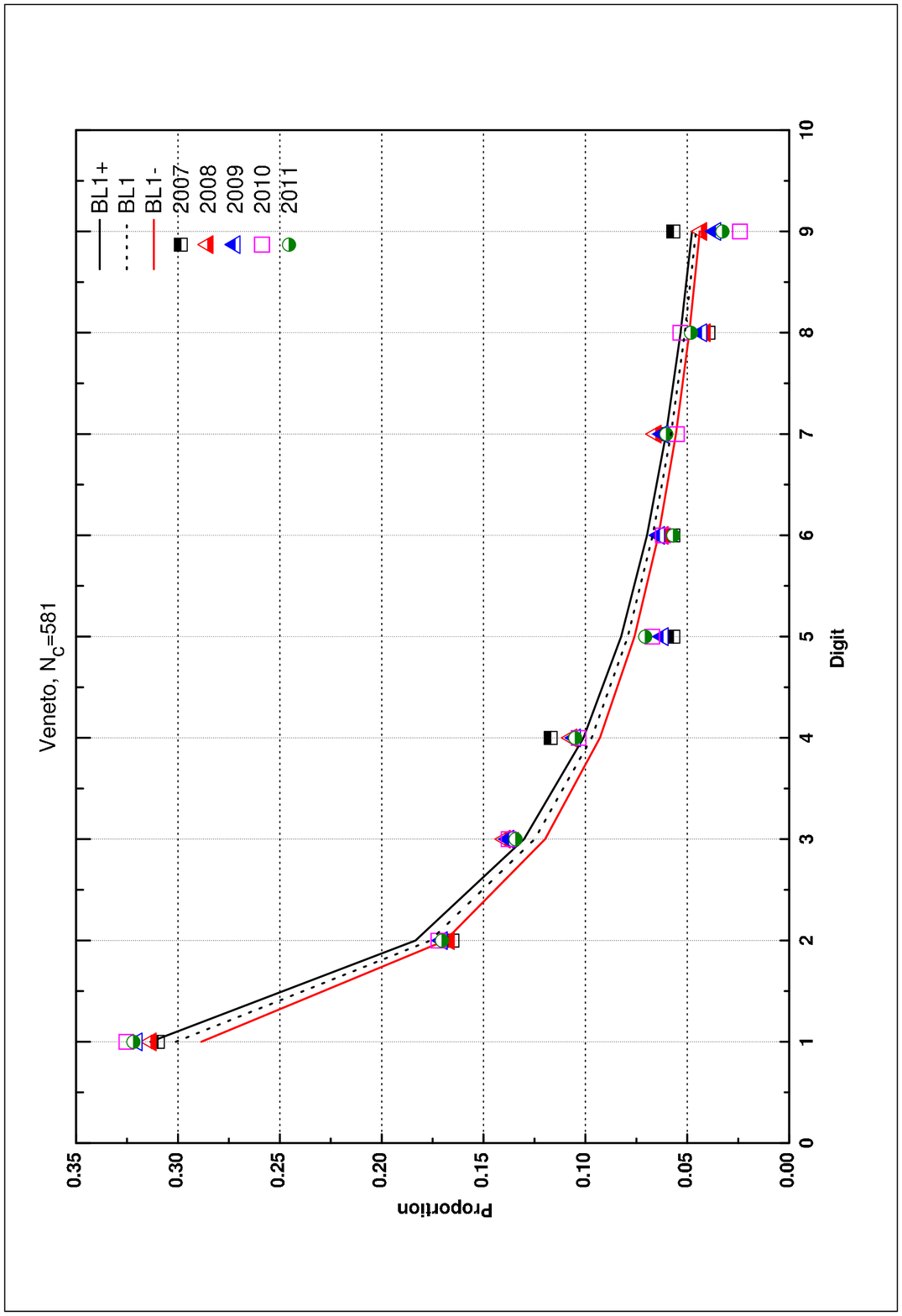}
\vspace*{50pt}  
  \caption{Veneto..}
  \label{fig:VENETO}
\hspace*{-45pt}
  \includegraphics[width=.77\linewidth, angle=270]{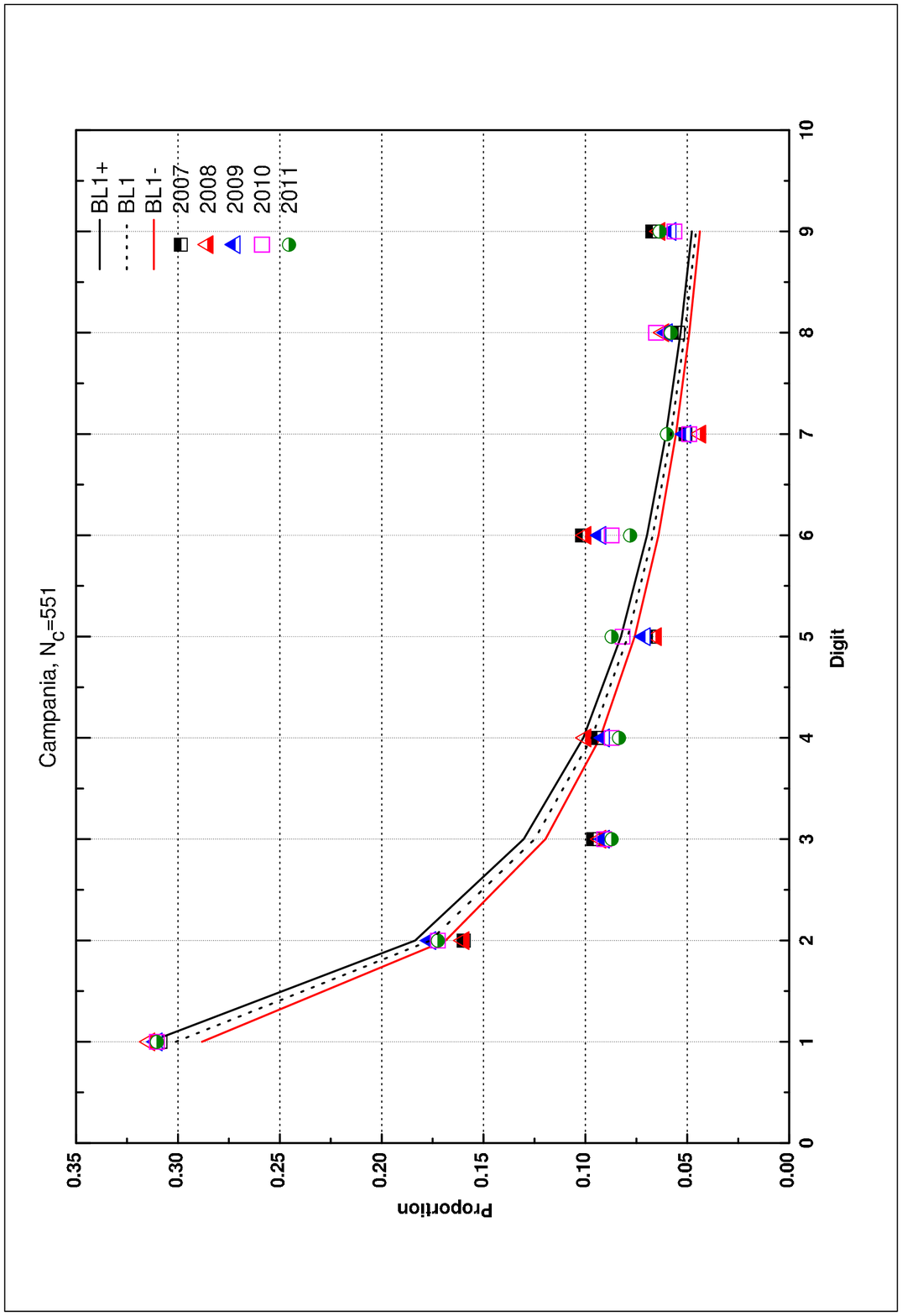}
\vspace*{-10pt}
  \caption{Campania.}
  \label{fig:CAMPANIA}
  %\end{center}
\end{figure}

%\begin{figure}
%\begin{center}
%\hspace*{-45pt}
%\vspace*{-10pt}
%  \includegraphics[width=.8\linewidth, angle=270]{FIGURE2.eps}
%\vspace*{5pt}
%  \caption{Piemonte.}
%  \label{fig:PIEMONTE}
%\end{center}
%\end{figure}

%\begin{figure}
%\begin{center}
%\hspace*{-45pt}
%\vspace*{-10pt}
 % \includegraphics[width=.8\linewidth, angle=270]{FIGURE3.eps}
 % \vspace*{5pt}
 % \caption{Veneto.}
 % \label{fig:VENETO}
%\end{center}
%\end{figure}

%\begin{figure}
%\begin{center}
%\hspace*{-45pt}
%\vspace*{-10pt}
%  \includegraphics[width=.8\linewidth, angle=270]{FIGURE4.eps}
%  \vspace*{5pt}
%  \caption{Campania.}
%  \label{fig:CAMPANIA}
%\end{center}
%\end{figure}
%\begin{figure}
%\begin{center}
%\hspace*{-45pt}
%\vspace*{-10pt}
 % \includegraphics[width=.8\linewidth, angle=270]{FIGURE5.eps}
%\vspace*{5pt}
%  \caption{Calabria.}
 % \label{fig:CALABRIA}
%\end{center}
%\end{figure}

\begin{figure}[t]
\hspace*{-45pt}
\vspace*{-65pt}
  \includegraphics[width=.77\linewidth, angle=270]{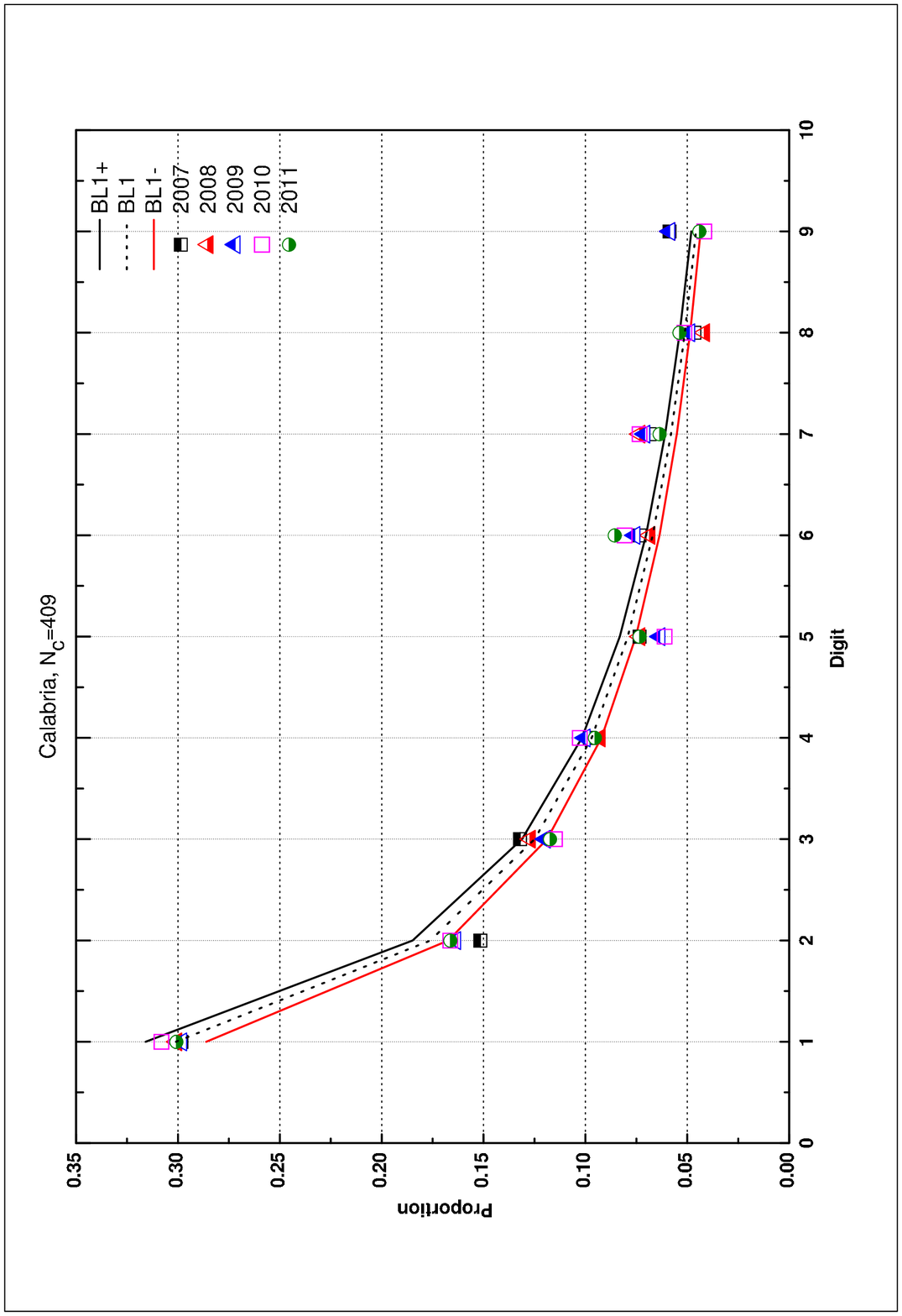}
\vspace*{50pt}  
  \caption{Calabria.}
 \label{fig:CALABRIA}
\hspace*{-45pt}
  \includegraphics[width=.77\linewidth, angle=270]{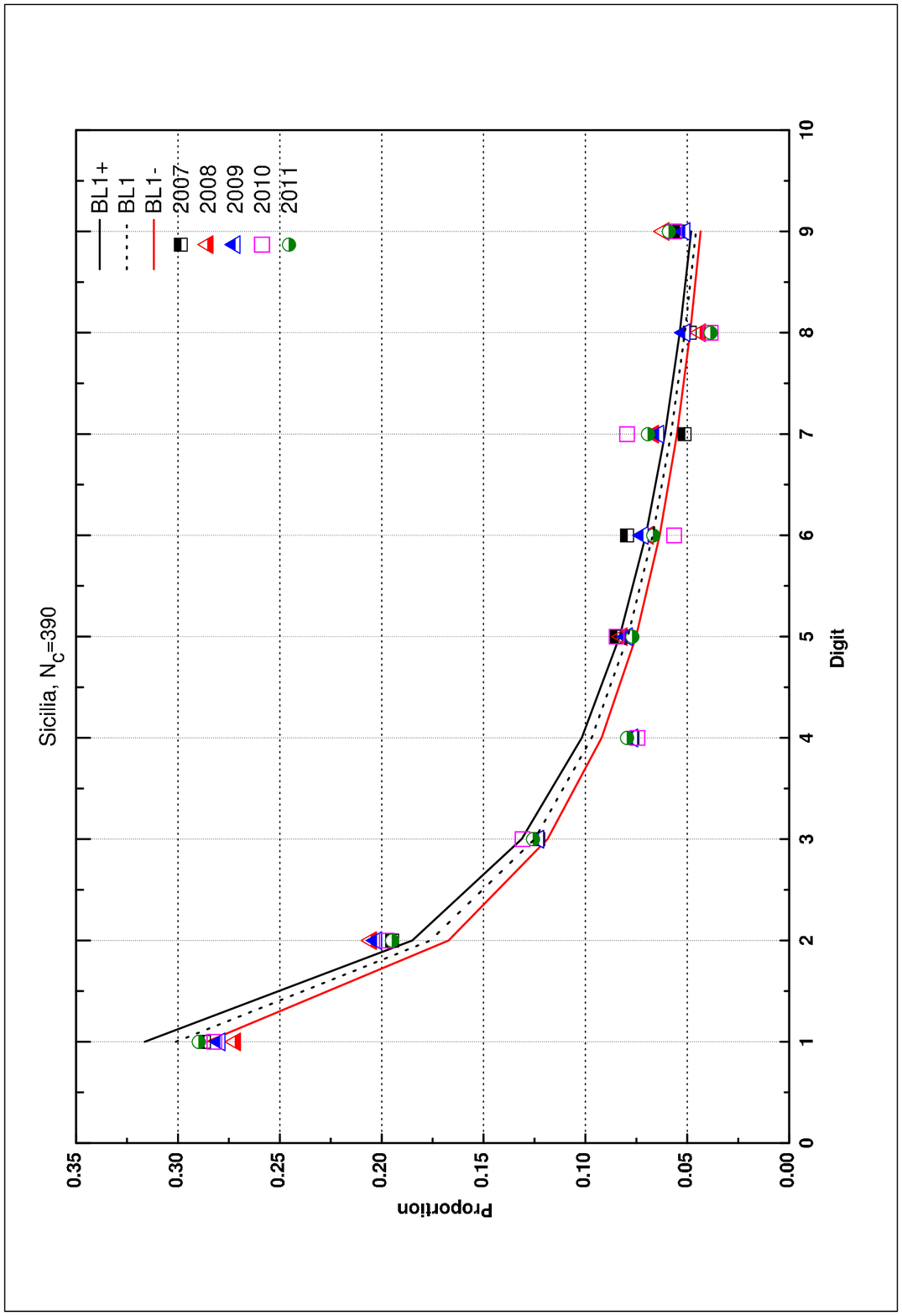}
\vspace*{-10pt}
  \caption{Sicilia.}
 \label{fig:SICILIA}
  %\end{center}
\end{figure}

%\begin{figure}
%\begin{center}
%\hspace*{-45pt}
%\vspace*{-10pt}
 % \includegraphics[width=.8\linewidth, angle=270]{FIGURE6.eps}
%\vspace*{5pt}
%  \caption{Sicilia.}
%  \label{fig:SICILIA}
%\end{center}
%\end{figure}

\begin{figure}[t]
\hspace*{-45pt}
\vspace*{-65pt}
  \includegraphics[width=.77\linewidth, angle=270]{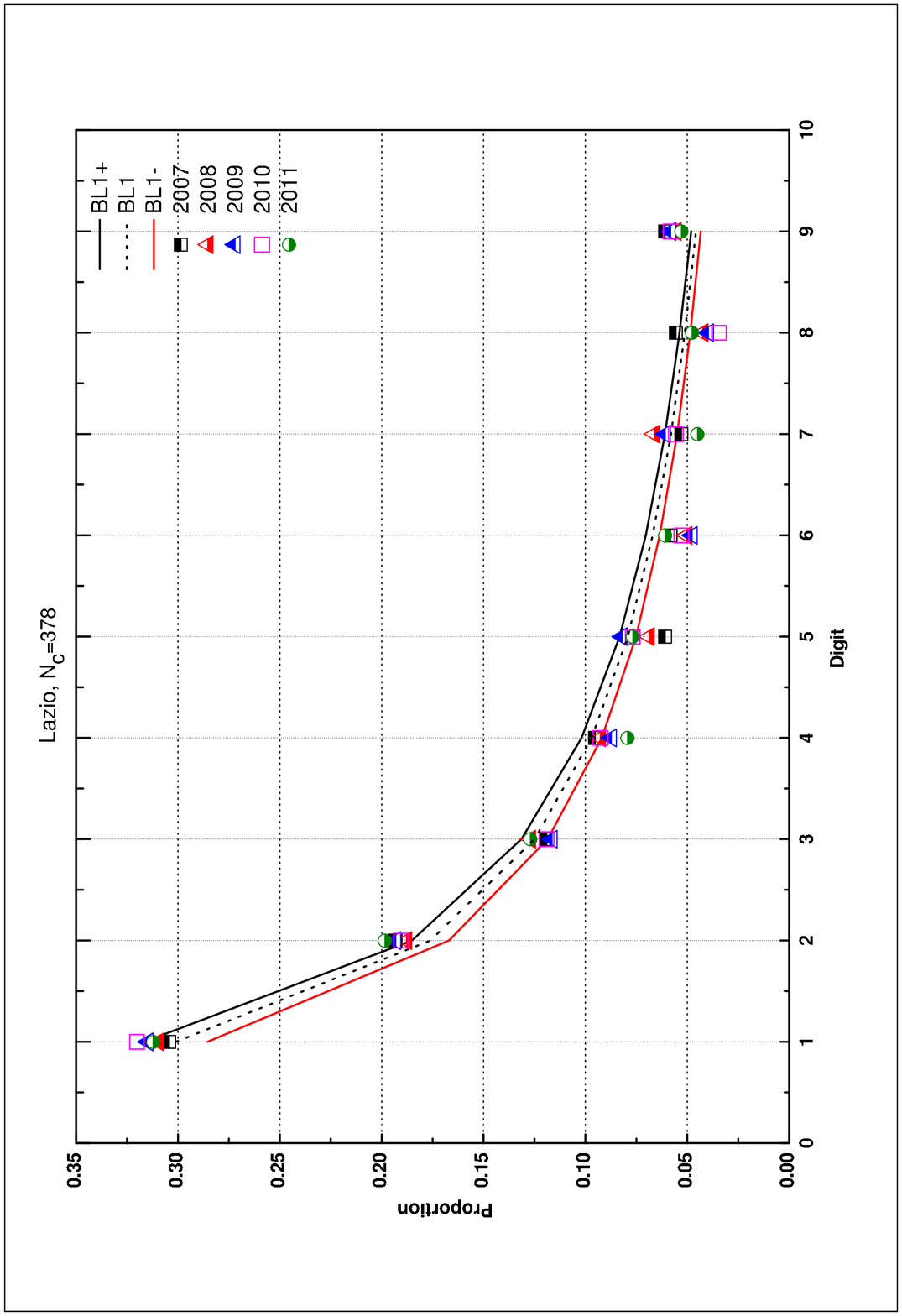}
\vspace*{50pt}  
  \caption{Lazio.}
 \label{fig:LAZIO}
\hspace*{-45pt}
  \includegraphics[width=.77\linewidth, angle=270]{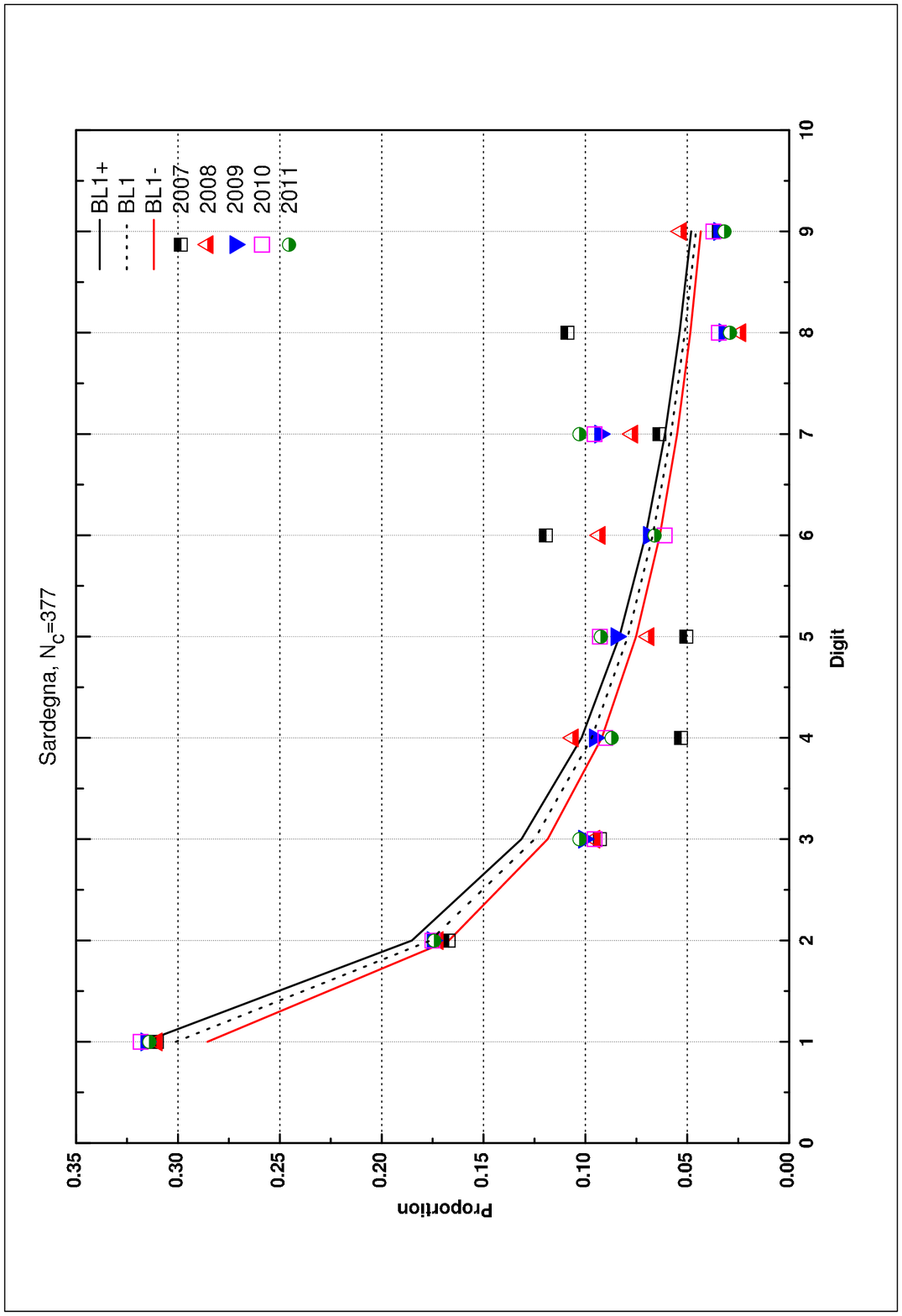}
\vspace*{-10pt}
  \caption{Sardegna.}
 \label{fig:SARDEGNA}
  %\end{center}
\end{figure}

%\begin{figure}
%\begin{center}
%\hspace*{-45pt}
%\vspace*{-10pt}
%  \includegraphics[width=.8\linewidth, angle=270]{FIGURE7.eps}
%\vspace*{5pt}
%  \caption{Lazio.}
%  \label{fig:LAZIO}
%\end{center}
%\end{figure}

%\begin{figure}
%\begin{center}
%\hspace*{-45pt}
%\vspace*{-10pt}
%  \includegraphics[width=.8\linewidth, angle=270]{FIGURE8.eps}
%\vspace*{5pt}
%  \caption{Sardegna.}
%  \label{fig:SARDEGNA}
%\end{center}
%\end{figure}

\begin{figure}[t]
\hspace*{-45pt}
\vspace*{-65pt}
  \includegraphics[width=.77\linewidth, angle=270]{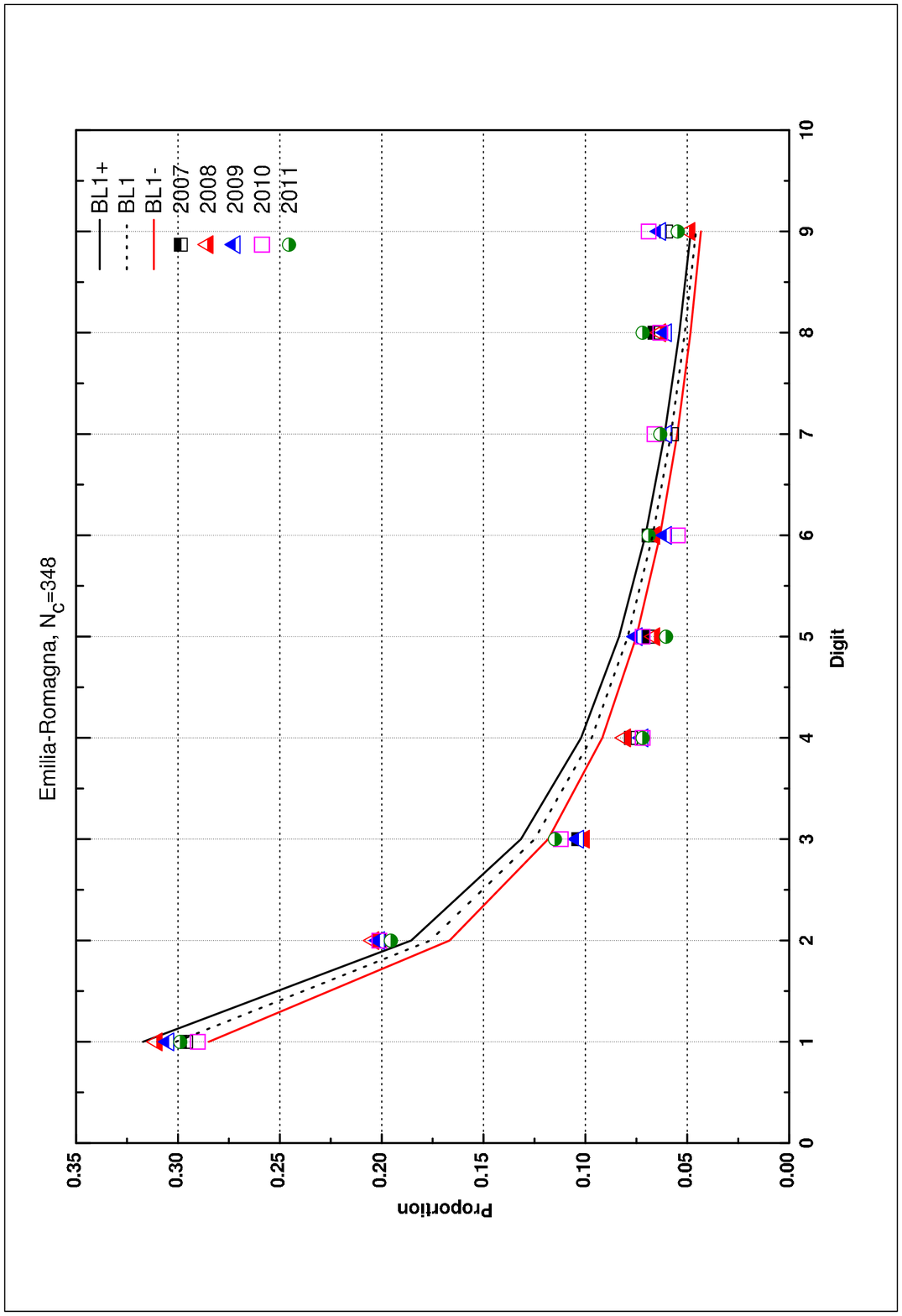}
\vspace*{50pt}  
  \caption{Emilia-Romagna.}
 \label{fig:EMILIA-ROMAGNA}
\hspace*{-45pt}
  \includegraphics[width=.77\linewidth, angle=270]{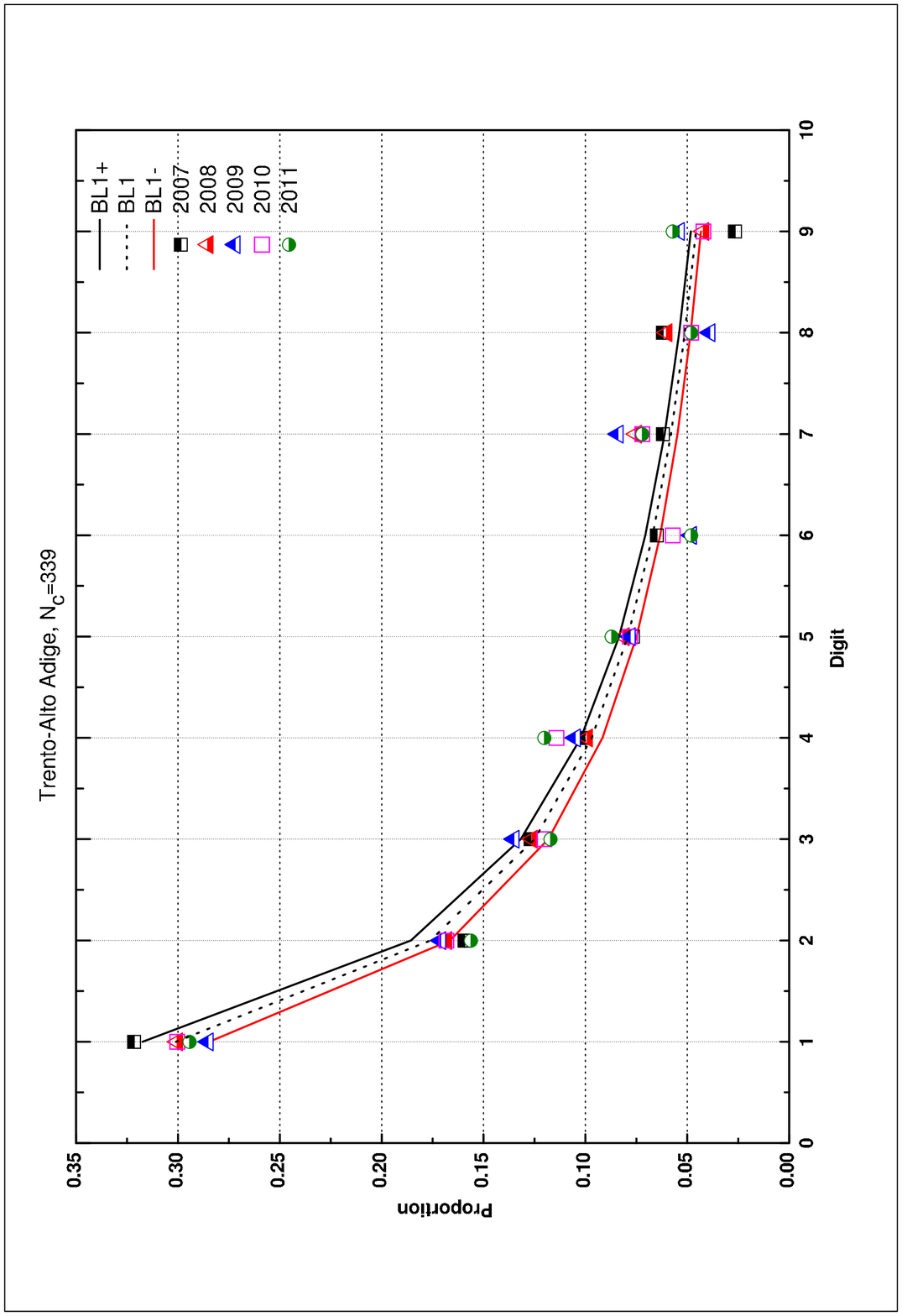}
\vspace*{-10pt}
  \caption{Trento-Alto Adige.}
 \label{fig:TRENTO-ALTO ADIGE}
  %\end{center}
\end{figure}

%\begin{figure}
%\begin{center}
%\hspace*{-45pt}
%\vspace*{-10pt}
%  \includegraphics[width=.8\linewidth, angle=270]{FIGURE9.eps}
%  \vspace*{5pt}
%  \caption{Emilia-Romagna.}
%  \label{fig:EMILIA-ROMAGNA}
%\end{center}
%\end{figure}

%\begin{figure}
%\begin{center}
%\hspace*{-45pt}
%\vspace*{-10pt}
%  \includegraphics[width=.8\linewidth, angle=270]{FIGURE10.eps}
%\vspace*{5pt}
%  \caption{Trento-Alto Adige.}
%  \label{fig:TRENTO-ALTO ADIGE}
%\end{center}
%\end{figure}

\begin{figure}[t]
\hspace*{-45pt}
\vspace*{-65pt}
  \includegraphics[width=.77\linewidth, angle=270]{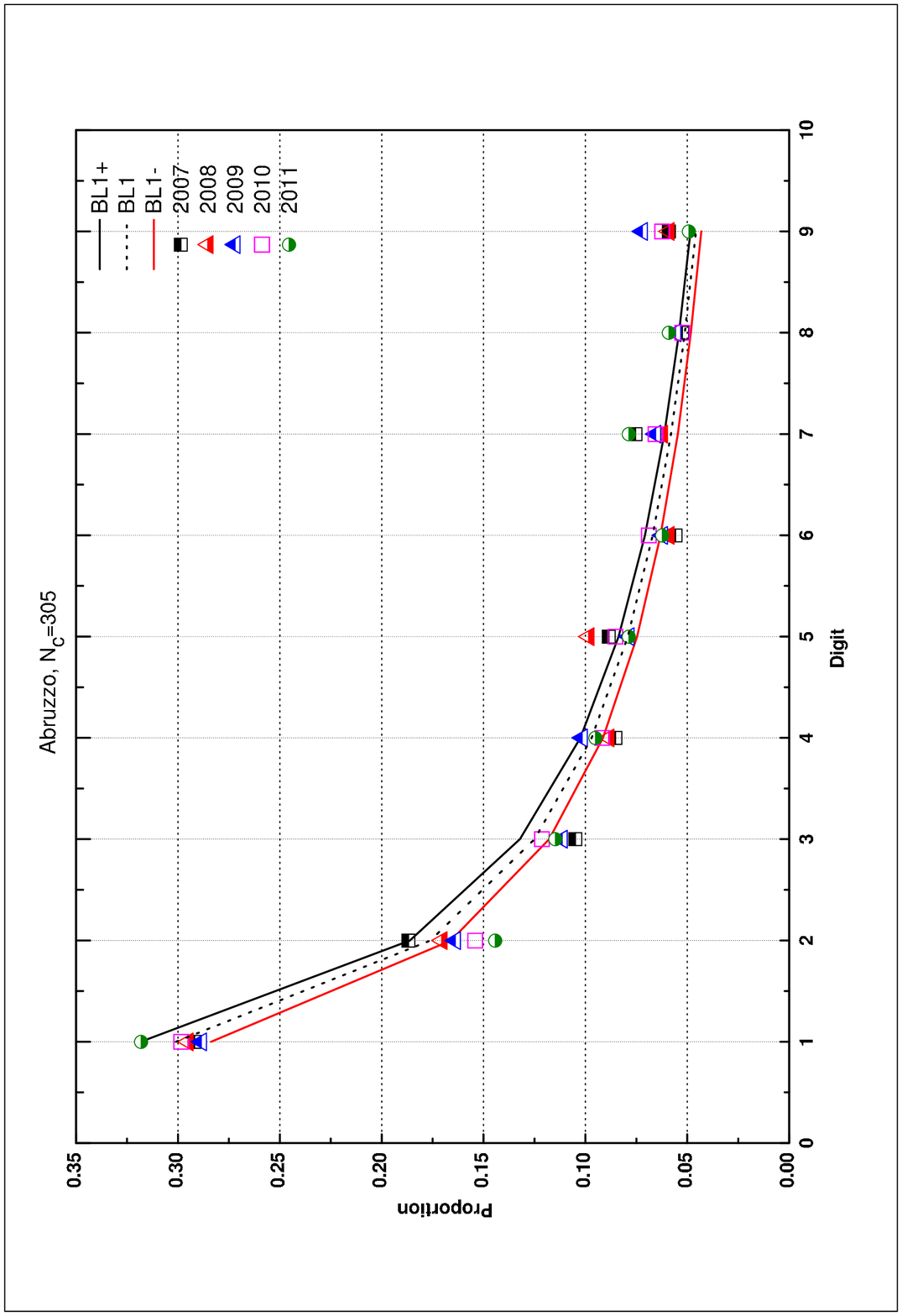}
\vspace*{50pt}  
  \caption{Abruzzo.}
 \label{fig:ABRUZZO}
\hspace*{-45pt}
  \includegraphics[width=.77\linewidth, angle=270]{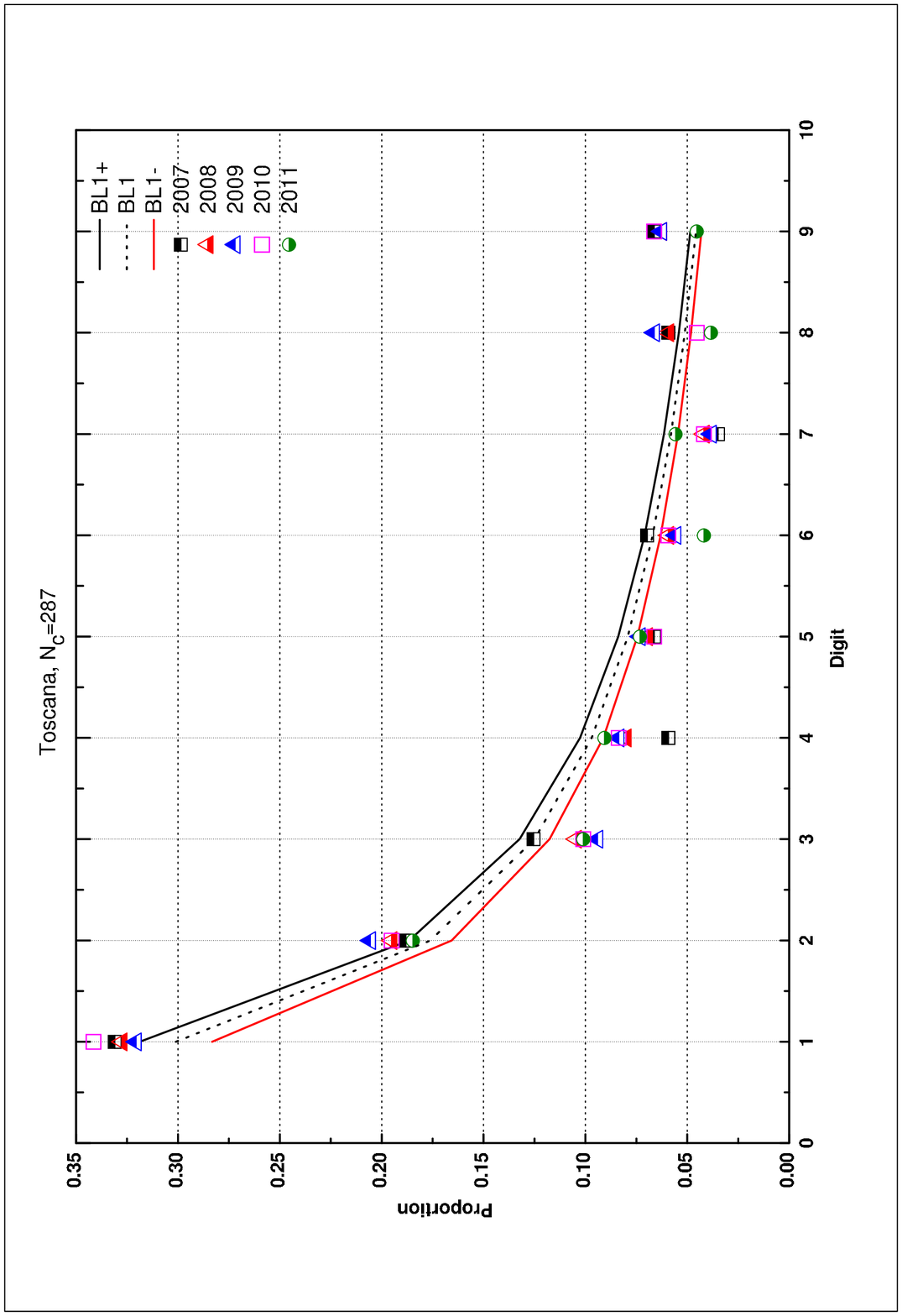}
\vspace*{-10pt}
  \caption{Toscana.}
 \label{fig:TOSCANA}
  %\end{center}
\end{figure}

%\begin{figure}
%\begin{center}
%\hspace*{-45pt}
%\vspace*{-10pt}
%  \includegraphics[width=.8\linewidth, angle=270]{FIGURE11.eps}
%\vspace*{5pt}
%  \caption{Abruzzo.}
%  \label{fig:ABRUZZO}
%\end{center}
%\end{figure}
%\begin{figure}
%\begin{center}
%\hspace*{-45pt}
%\vspace*{-10pt}
%  \includegraphics[width=.8\linewidth, angle=270]{FIGURE12.eps}
%\vspace*{5pt}
%  \caption{Toscana.}
%  \label{fig:TOSCANA}
%\end{center}
%\end{figure}

\begin{figure}[t]
\hspace*{-45pt}
\vspace*{-65pt}
  \includegraphics[width=.77\linewidth, angle=270]{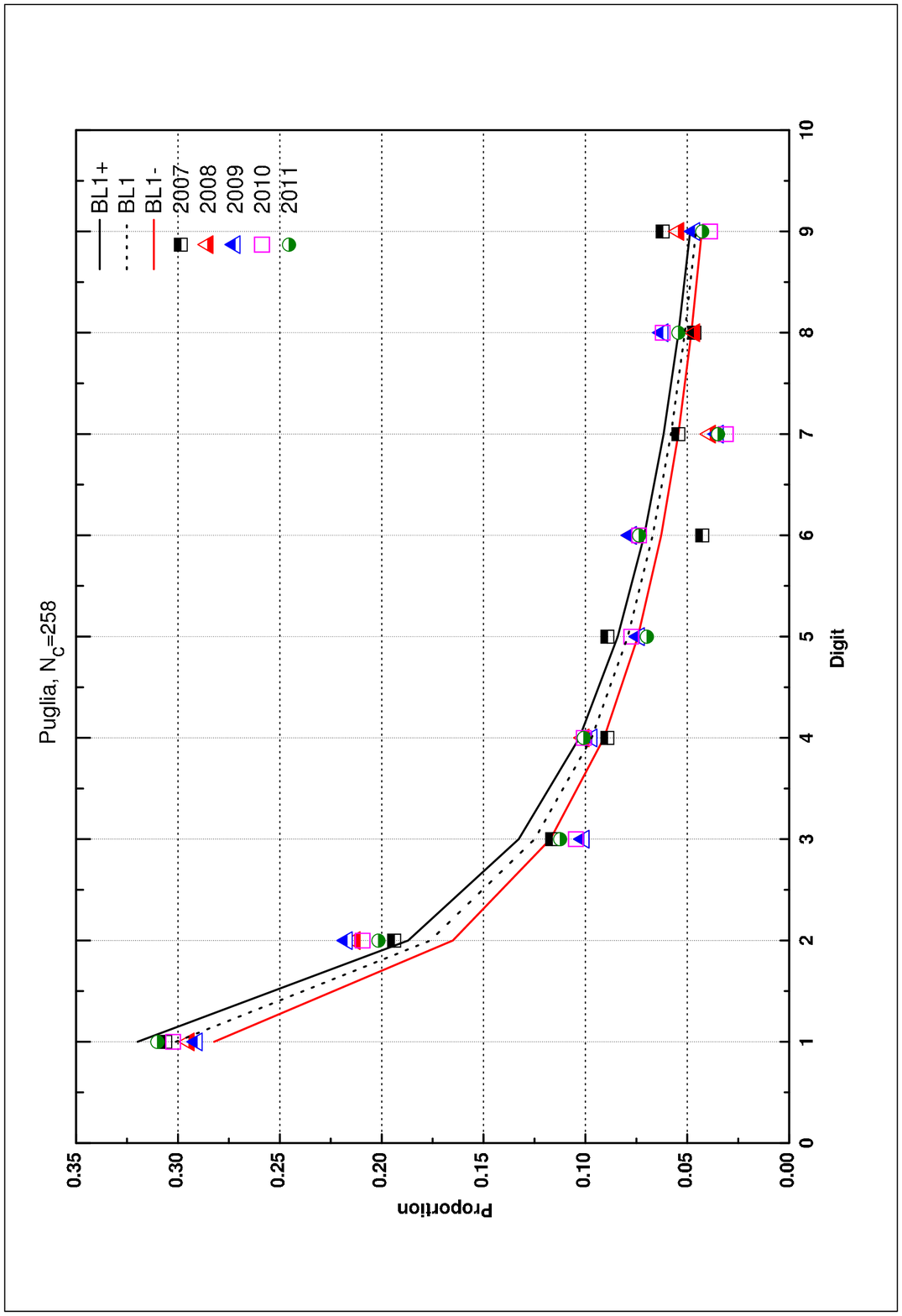}
\vspace*{50pt}  
  \caption{Puglia.}
 \label{fig:PUGLIA}
\hspace*{-45pt}
  \includegraphics[width=.77\linewidth, angle=270]{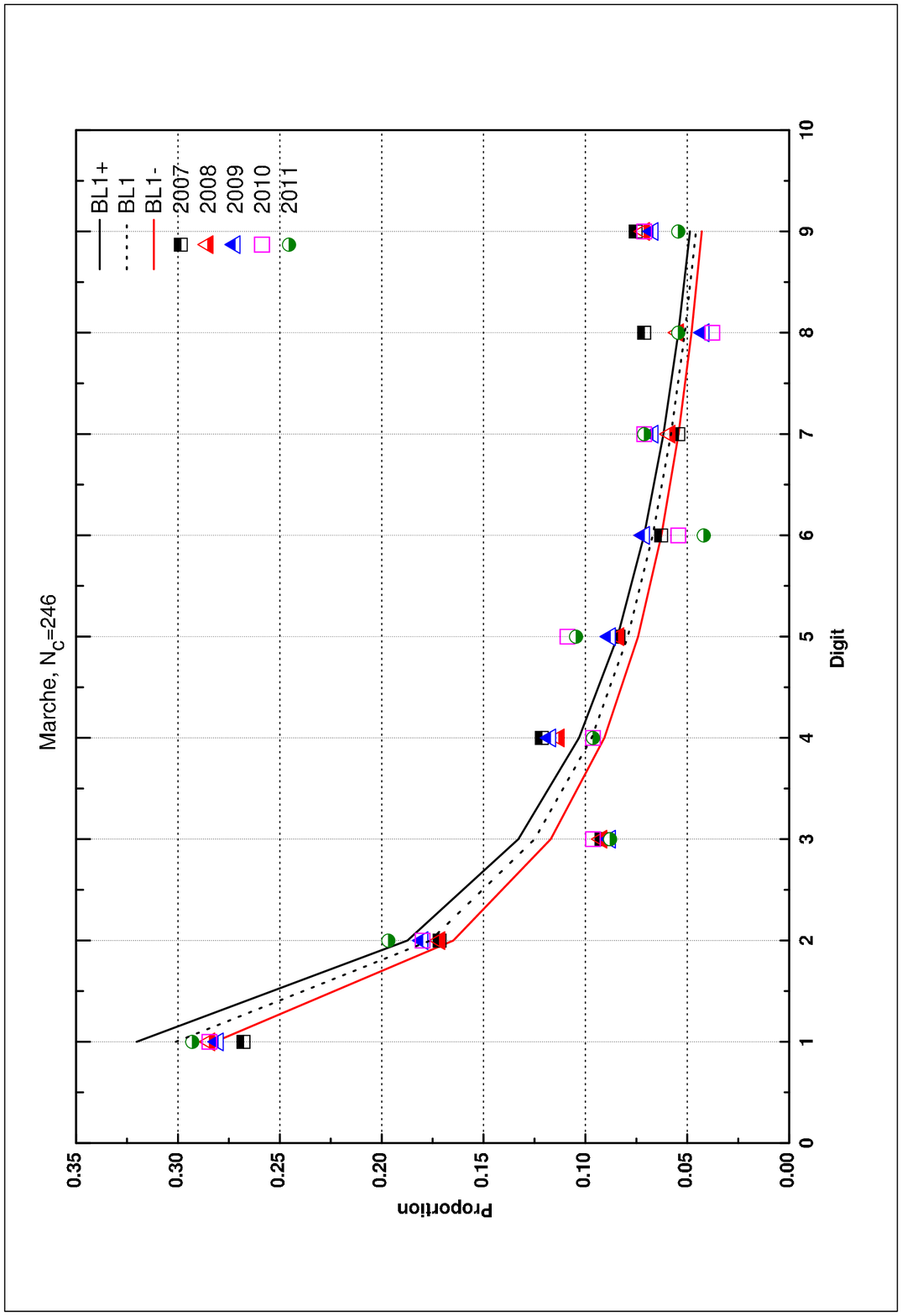}
\vspace*{-10pt}
  \caption{Marche.}
 \label{fig:MARCHE}
  %\end{center}
\end{figure}

%\begin{figure}
%\begin{center}
%\hspace*{-45pt}
%\vspace*{-10pt}
%  \includegraphics[width=.8\linewidth, angle=270]{FIGURE13.eps}
%\vspace*{5pt}
%  \caption{Puglia.}
%  \label{fig:PUGLIA}
%\end{center}
%\end{figure}

%\begin{figure}
%\begin{center}
%\hspace*{-45pt}
%\vspace*{-10pt}
%  \includegraphics[width=.8\linewidth, angle=270]{FIGURE14.eps}
%\vspace*{5pt}
%  \caption{Marche.}
%  \label{fig:MARCHE}
%\end{center}
%\end{figure}

%\begin{figure}
%\begin{center}
%\hspace*{-45pt}
%\vspace*{-10pt}
%  \includegraphics[width=.8\linewidth, angle=270]{FIGURE15.eps}
%\vspace*{5pt}
%  \caption{Liguria.}
%  \label{fig:LIGURIA}
%\end{center}
%\end{figure}

\begin{figure}[t]
\hspace*{-45pt}
\vspace*{-65pt}
  \includegraphics[width=.77\linewidth, angle=270]{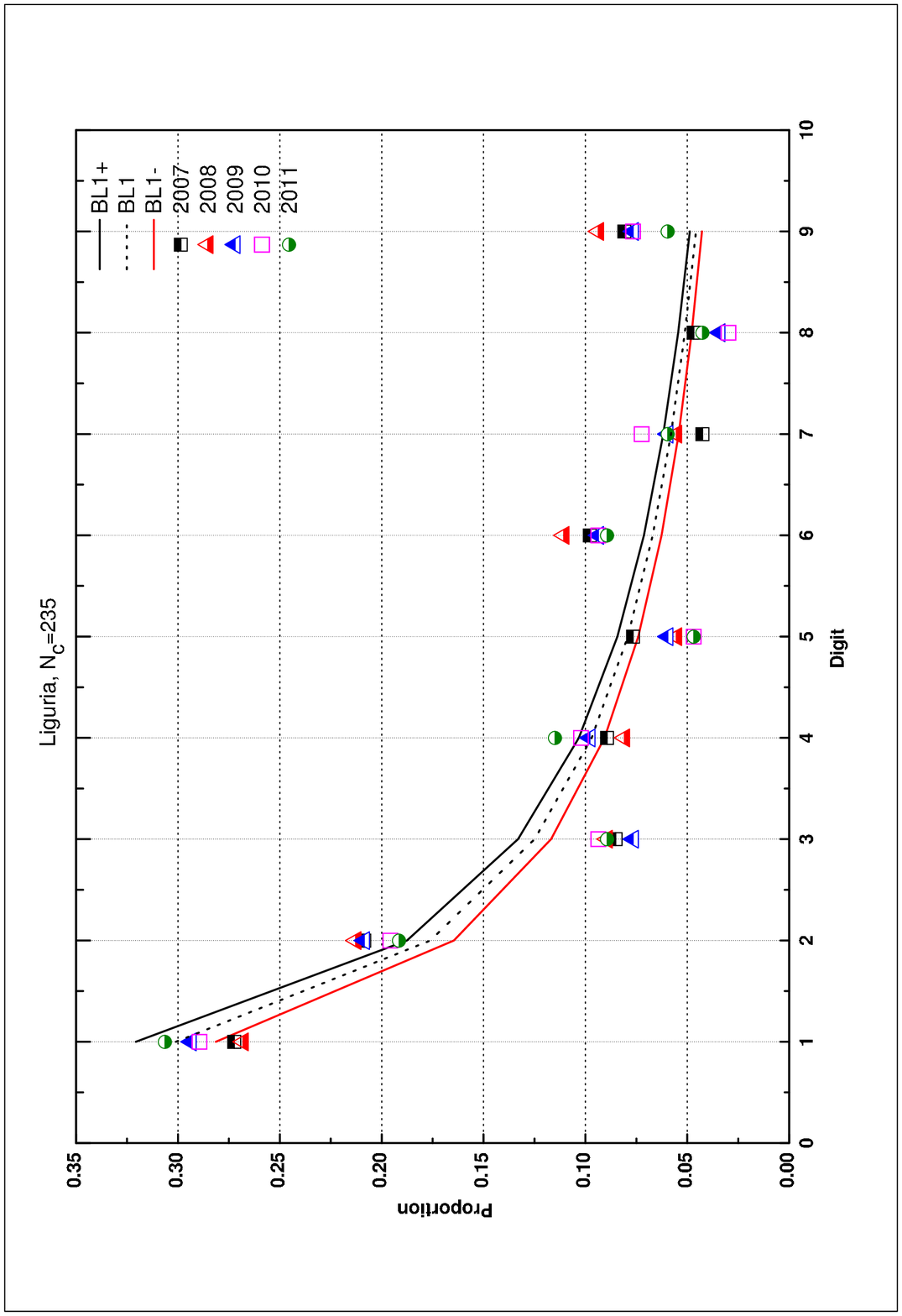}
\vspace*{50pt}  
  \caption{Liguria.}
 \label{fig:LIGURIA}
\hspace*{-45pt}
  \includegraphics[width=.77\linewidth, angle=270]{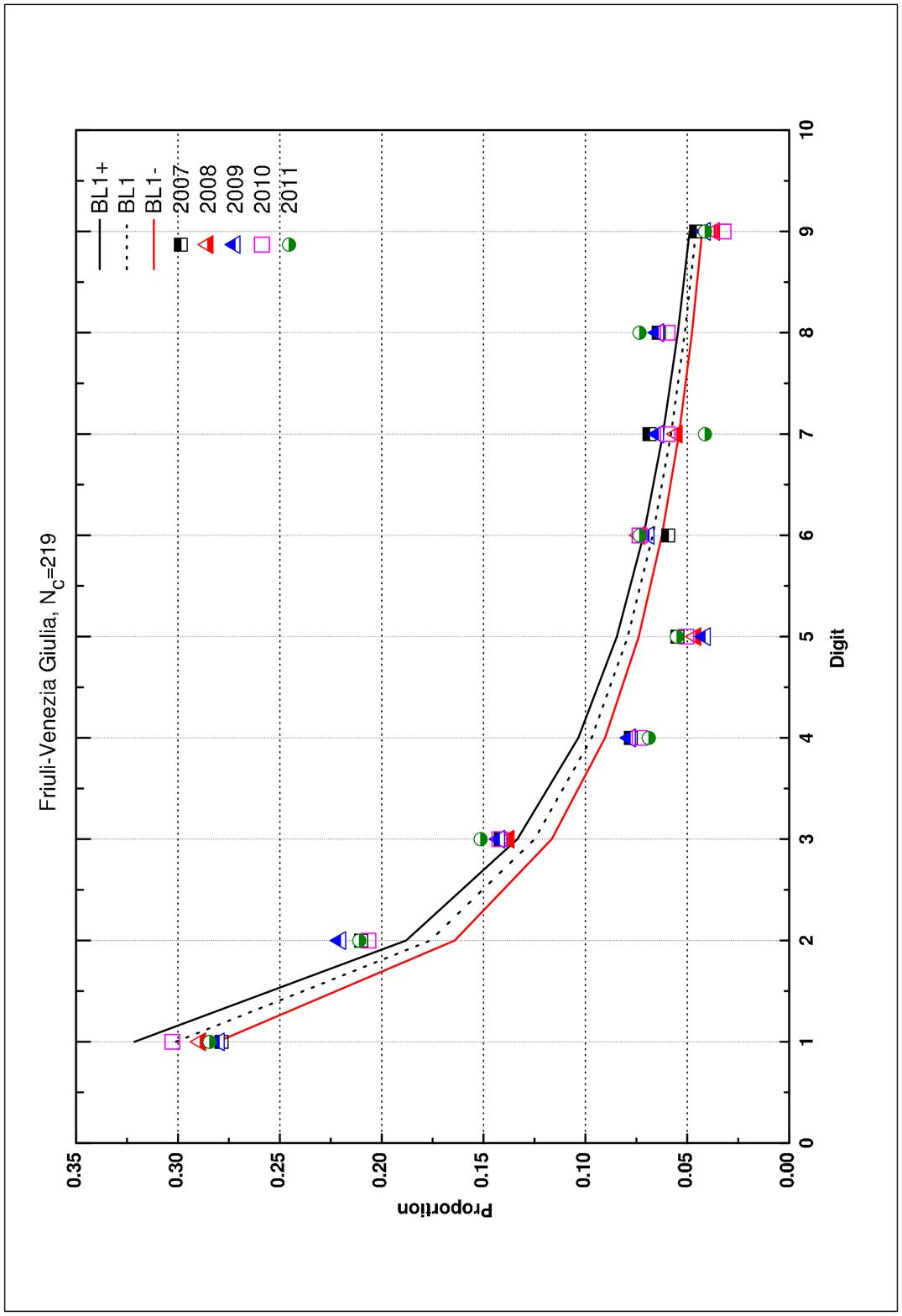}
\vspace*{-10pt}
  \caption{Friuli-Venezia Giulia.}
 \label{fig:FRIULI-VENEZIA GIULIA}
  %\end{center}
\end{figure}

%\begin{figure}
%\begin{center}
%\hspace*{-45pt}
%\vspace*{-10pt}
%  \includegraphics[width=.8\linewidth, angle=270]{FIGURE16.eps}
%\vspace*{5pt}
%  \caption{Friuli-Venezia Giulia.}
%  \label{fig:FRIULI-VENEZIA Giulia}
%\end{center}
%\end{figure}

%\begin{figure}
%\begin{center}
%\hspace*{-45pt}
%\vspace*{-10pt}
%  \includegraphics[width=.8\linewidth, angle=270]{FIGURE17.eps}
%\vspace*{5pt}
%  \caption{Molise.}
%  \label{fig:MOLISE}
%\end{center}
%\end{figure}

\begin{figure}[t]
\hspace*{-45pt}
\vspace*{-65pt}
  \includegraphics[width=.77\linewidth, angle=270]{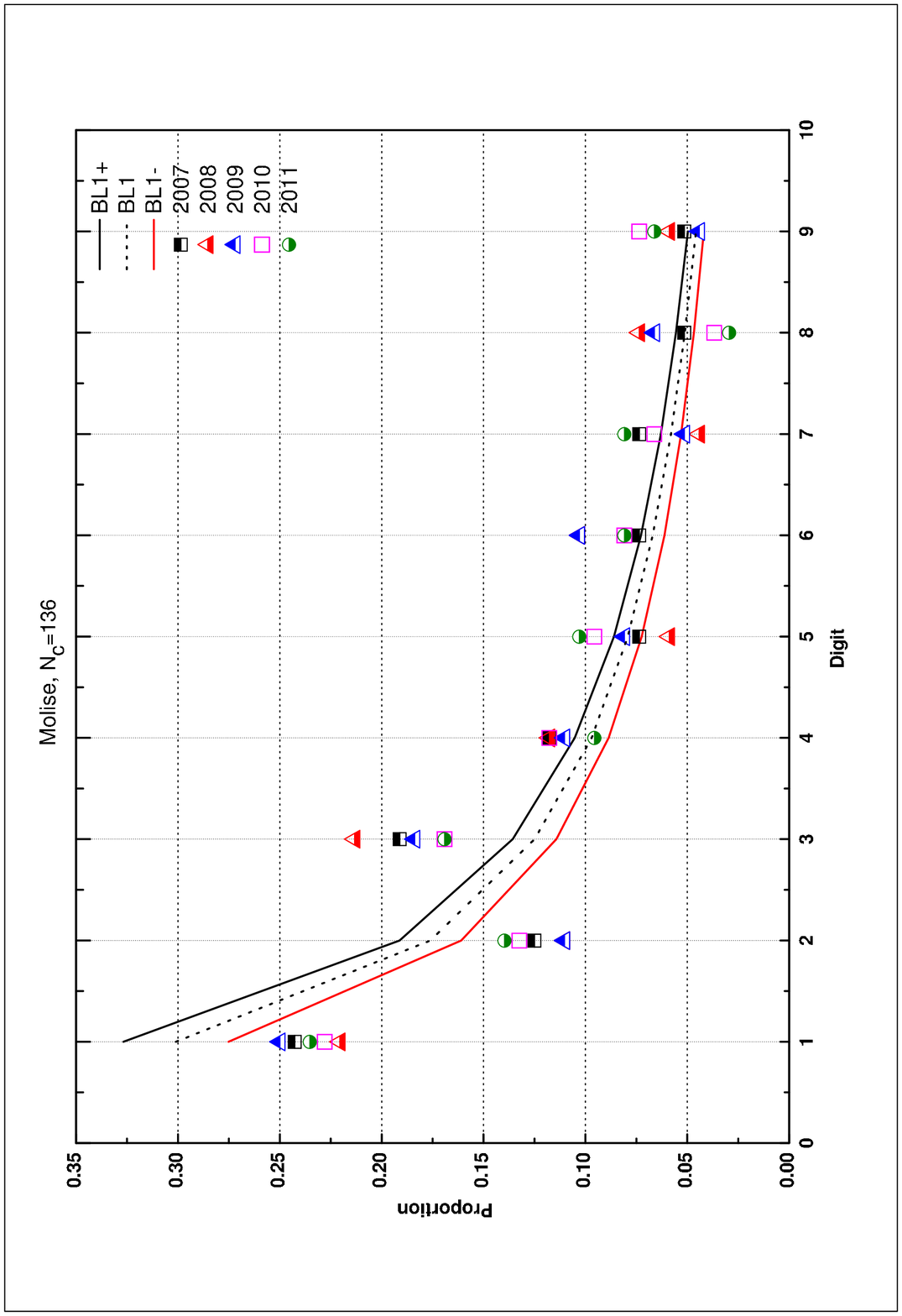}
\vspace*{50pt}  
  \caption{Molise.}
 \label{fig:MOLISE}
\hspace*{-45pt}
  \includegraphics[width=.77\linewidth, angle=270]{FIGURE1.eps}
\vspace*{-10pt}
  \caption{Basilicata.}
 \label{fig:BASILICATA}
  %\end{center}
\end{figure}

%\begin{figure}
%\begin{center}
%\hspace*{-45pt}
%\vspace*{-10pt}
%  \includegraphics[width=.8\linewidth, angle=270]{FIGURE18.eps}
%\vspace*{5pt}
%  \caption{Basilicata.}
%  \label{fig:BASILICATA}
%\end{center}
%\end{figure}

%\begin{figure}
%\begin{center}
%\hspace*{-45pt}
%\vspace*{-10pt}
%  \includegraphics[width=.8\linewidth, angle=270]{FIGURE19.eps}
%\vspace*{5pt}
 % \caption{Umbria.}
 % \label{fig:UMBRIA}
%\end{center}
%\end{figure}

\begin{figure}[t]
\hspace*{-45pt}
\vspace*{-65pt}
  \includegraphics[width=.77\linewidth, angle=270]{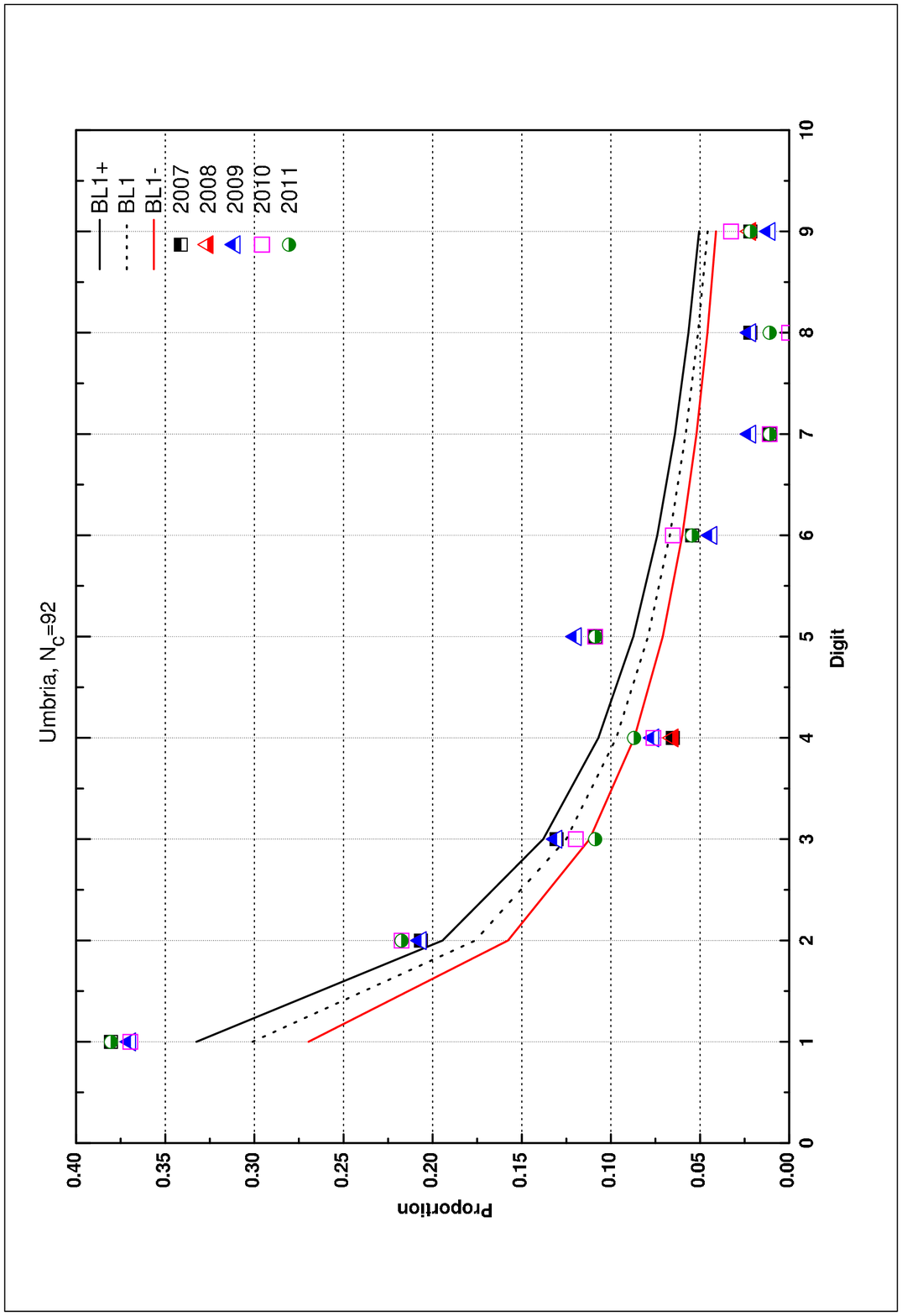}
\vspace*{50pt}  
  \caption{Umbria.}
 \label{fig:UMBRIA}
\hspace*{-45pt}
  \includegraphics[width=.77\linewidth, angle=270]{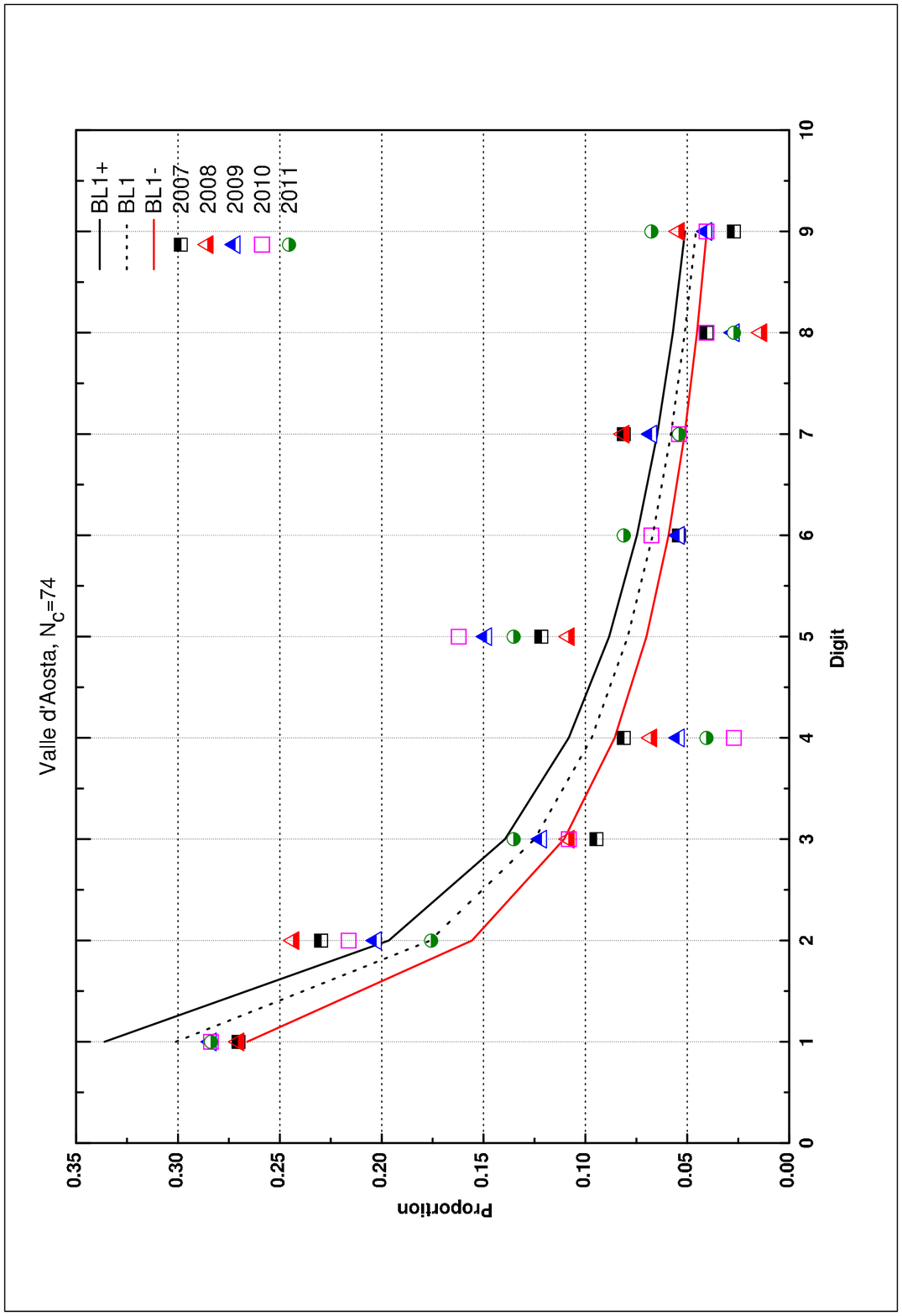}
\vspace*{-10pt}
  \caption{Valle d'Aosta.}
 \label{fig:VALLE d'AOSTA}
  %\end{center}
\end{figure}

%\begin{figure}
%\begin{center}
%\hspace*{-45pt}
%\vspace*{-10pt}
%  \includegraphics[width=.8\linewidth, angle=270]{FIGURE20.eps}
%\vspace*{5pt}
%  \caption{Valle d'Aosta.}
%  \label{fig:VALLE d'AOSTA}
%\end{center}
%\end{figure}

%\begin{center}
%\textbf{Caption:} Lombardia
%\end{center}
%\begin{center}
%INSERT FIGURE 2 ABOUT HERE\\
%\textbf{Caption:} Piemonte
%\end{center}
%\begin{center}
%INSERT FIGURE 3 ABOUT HERE\\
%\textbf{Caption:} Veneto
%\end{center}
%\begin{center}
%INSERT FIGURE 4 ABOUT HERE\\
%\textbf{Caption:} Campania
%\end{center}

\begin{figure}[t]
\vspace*{-80pt}
\hspace*{-45pt}
\vspace*{-65pt}
  \includegraphics[width=1.0\linewidth, angle=270]{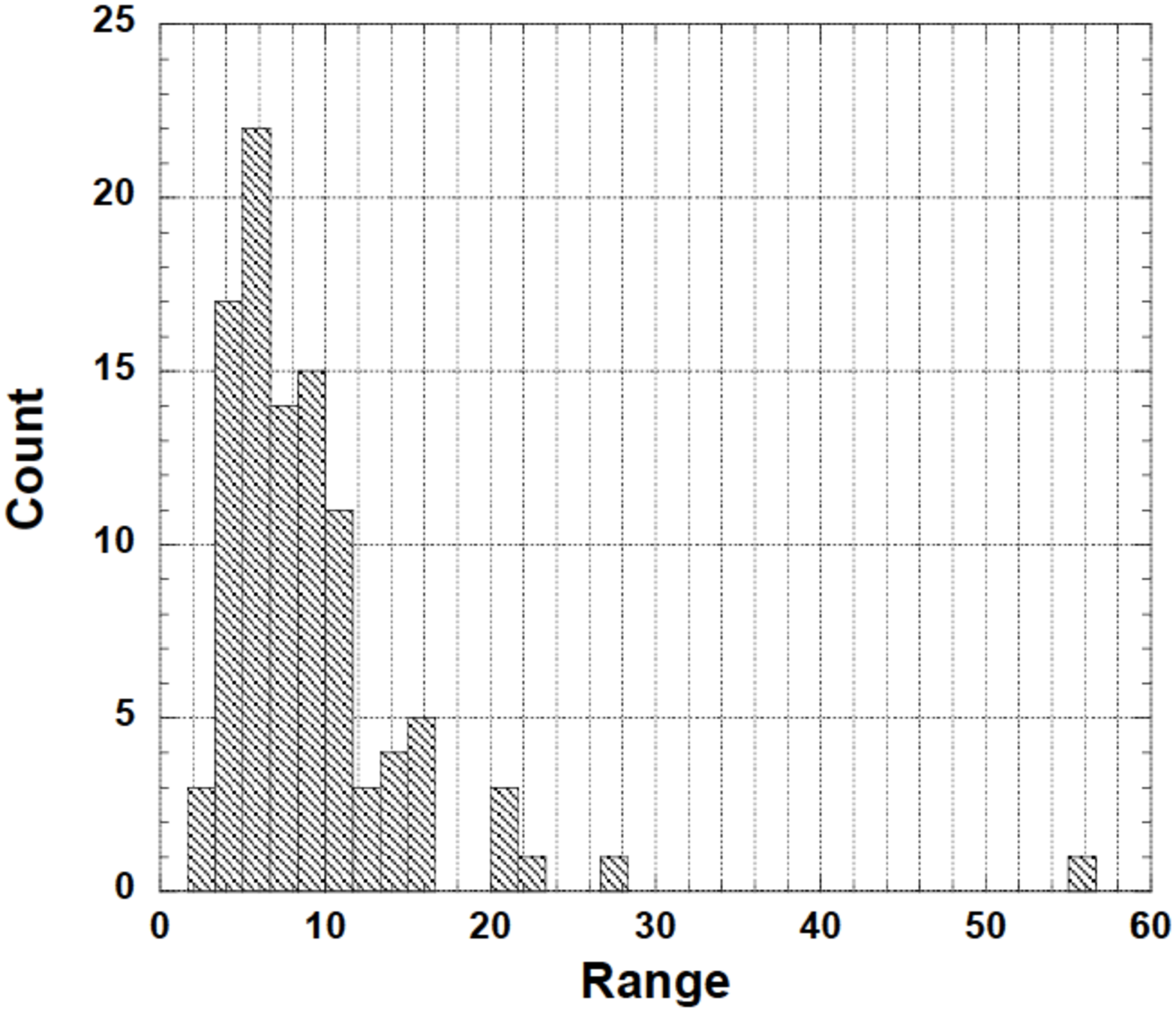}
\vspace*{50pt}  
  \caption{Distribution of  yearly $\chi^2$ values in BL1
analysis of  AIT in the 20 IT  regions.}
 \label{fig:DISTRi}
\hspace*{-45pt}
  \includegraphics[width=1.0\linewidth, angle=270]{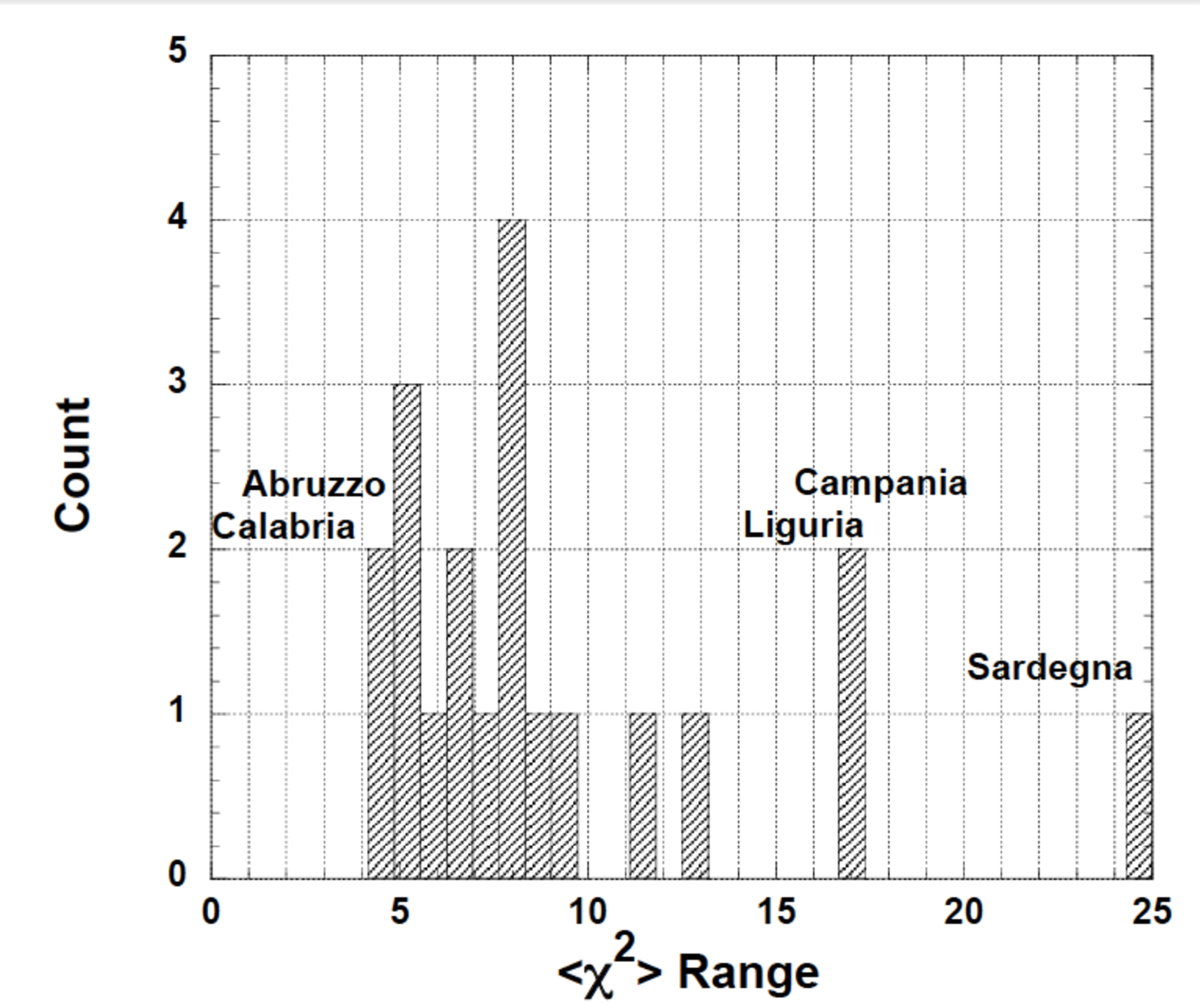}
\vspace*{-10pt}
  \caption{Distribution of  the mean (average over the
quinquennium) $\chi^2$
 values in BL1 analysis of  AIT in the 20 IT  regions.}
 \label{fig:VALLE d'AOSTA}
  %\end{center}
\end{figure}

%\begin{center}
%\textbf{Caption:} Distribution of  yearly $\chi^2$ values in BL1
%analysis of  AIT in the 20 IT  regions.
% \end{center}
%\begin{center}
%INSERT FIGURE 22 ABOUT HERE\\
%\textbf{Caption:} Distribution of  the mean (average over the
%quinquennium) $\chi^2$
% values in BL1 analysis of  AIT in the 20 IT  regions.
%\end{center}

\clearpage
\begin{table} \begin{center}
\vspace{-50pt}
\begin{tabular}[t]{|c|c|c|c|c|c|c|c|c|c|c|}
  \hline Digit  &   1   &   2   &   3   &   4   &   5   &   6   &   7   &   8   &   9   &$ \chi^2 $\\ \hline\hline
Year&\multicolumn{9}{|c|}{  Lombardia,  $N_c = $ 1546  for 2007-2019  ;  $N_c = $ 1544  for 2010-2011 } &                                                                        % \\ \hline Year/Digit &   1.000   &   2.000   &   3.000   &   4.000   &   5.000   &   6.000   &   7.000   &   8.000   &   9.000   &       &$ \chi^2 $  \\ \hline
  \\ \hline2007     &   0.298   &   0.168   &   0.118   &   0.092   &   0.087   &   0.074   &   0.064   &   0.050   &   0.048   &   5.297
 \\ \hline2008  &   0.297   &   0.175   &   0.115   &   0.089   &   0.085   &   0.073   &   0.070   &   0.045   &   0.051   &   9.692
 \\ \hline2009  &   0.299   &   0.176   &   0.114   &   0.093   &   0.085   &   0.072   &   0.070   &   0.048   &   0.043   &   7.348
 \\ \hline2010  &   0.296   &   0.172   &   0.119   &   0.096   &   0.082   &   0.071   &   0.069   &   0.054   &   0.043   &   4.674
 \\ \hline2011  &   0.300   &   0.172   &   0.117   &   0.097   &   0.078   &   0.074   &   0.067   &   0.051   &   0.043   &   4.836
 \\ \hline  &\multicolumn{9}{|c|}{Piemonte, $N_c = $ 1206 } &
 % \\ \hline Year/Digit &   1.000   &   2.000   &   3.000   &   4.000   &   5.000   &   6.000   &   7.000   &   8.000   &   9.000   &       &$ \chi^2 $\\ \hline
   \\  \hline2007   &   0.304   &   0.185   &   0.119   &   0.091   &   0.088   &   0.061   &   0.064   &   0.052   &   0.036   &   6.279
 \\ \hline2008  &   0.300   &   0.186   &   0.112   &   0.096   &   0.085   &   0.065   &   0.065   &   0.044   &   0.047   &   5.175
 \\ \hline2009  &   0.303   &   0.180   &   0.110   &   0.095   &   0.090   &   0.061   &   0.061   &   0.053   &   0.046   &   4.925
 \\ \hline2010  &   0.304   &   0.185   &   0.111   &   0.094   &   0.086   &   0.070   &   0.055   &   0.054   &   0.041   &   4.327
 \\ \hline2011  &   0.299   &   0.186   &   0.114   &   0.095   &   0.076   &   0.080   &   0.056   &   0.051   &   0.042   &   5.709
 \\ \hline &\multicolumn{9}{|c|}{Veneto,    $N_c = $ 581 } &
 % \\ \hline Year/Digit &   1.000   &   2.000   &   3.000   &   4.000   &   5.000   &   6.000   &   7.000   &   8.000   &   9.000   &       &$ \chi^2 $\\ \hline
    \\ \hline2007   &   0.310   &   0.165   &   0.136   &   0.117   &   0.057   &   0.057   &   0.062   &   0.040   &   0.057   &   11.327
 \\ \hline2008  &   0.313   &   0.167   &   0.139   &   0.107   &   0.062   &   0.062   &   0.065   &   0.041   &   0.043   &   6.251
 \\ \hline2009  &   0.320   &   0.170   &   0.138   &   0.105   &   0.062   &   0.064   &   0.062   &   0.043   &   0.036   &   6.309
 \\ \hline2010  &   0.325   &   0.172   &   0.138   &   0.103   &   0.067   &   0.062   &   0.055   &   0.053   &   0.024   &   9.569
 \\ \hline2011  &   0.322   &   0.170   &   0.134   &   0.105   &   0.071   &   0.057   &   0.060   &   0.048   &   0.033   &   5.492
 \\ \hline &\multicolumn{9}{|c|}{Campania,      $N_c = $ 551 } &
 % \\ \hline Year/Digit &   1.000   &   2.000   &   3.000   &   4.000   &   5.000   &   6.000   &   7.000   &   8.000   &   9.000   &       &$ \chi^2 $\\ \hline
  \\ \hline2007     &   0.309   &   0.160   &   0.096   &   0.094   &   0.067   &   0.102   &   0.051   &   0.054   &   0.067   &   \underline{21.647}
 \\ \hline2008  &   0.314   &   0.160   &   0.093   &   0.100   &   0.065   &   0.100   &   0.044   &   0.062   &   0.064   &   \underline{23.02}1
 \\ \hline2009  &   0.310   &   0.176   &   0.091   &   0.091   &   0.071   &   0.093   &   0.051   &   0.060   &   0.058   &   14.559
 \\ \hline2010  &   0.310   &   0.172   &   0.091   &   0.087   &   0.082   &   0.087   &   0.049   &   0.065   &   0.056   &   13.556
 \\ \hline2011  &   0.310   &   0.172   &   0.087   &   0.083   &   0.087   &   0.078   &   0.060   &   0.058   &   0.064   &   13.336
 \\ \hline &\multicolumn{9}{|c|}{Calabria,      $N_c = $ 409 } &
 % \\ \hline Year/Digit &   1.000   &   2.000   &   3.000   &   4.000   &   5.000   &   6.000   &   7.000   &   8.000   &   9.000   &       &$ \chi^2 $\\ \hline
 \\ \hline2007  &   0.298   &   0.152   &   0.132   &   0.098   &   0.073   &   0.073   &   0.068   &   0.046   &   0.059   &   4.439
 \\ \hline2008  &   0.301   &   0.164   &   0.127   &   0.093   &   0.073   &   0.068   &   0.073   &   0.042   &   0.059   &   4.514
 \\ \hline2009  &   0.298   &   0.164   &   0.120   &   0.100   &   0.064   &   0.076   &   0.071   &   0.049   &   0.059   &   4.939
 \\ \hline2010  &   0.308   &   0.166   &   0.115   &   0.103   &   0.061   &   0.081   &   0.073   &   0.051   &   0.042   &   5.420
 \\ \hline2011  &   0.301   &   0.166   &   0.117   &   0.095   &   0.073   &   0.086   &   0.064   &   0.054   &   0.044   &   3.020
 \\ \hline &\multicolumn{9}{|c|}{Sicilia,   $N_c = $ 390 } &
 % \\ \hline Year/Digit &   1.000   &   2.000   &   3.000   &   4.000   &   5.000   &   6.000   &   7.000   &   8.000   &   9.000   &       &$ \chi^2 $\\ \hline
 \\ \hline2007  &   0.287   &   0.195   &   0.123   &   0.077   &   0.085   &   0.079   &   0.051   &   0.049   &   0.054   &   4.614
 \\ \hline2008  &   0.272   &   0.205   &   0.123   &   0.077   &   0.082   &   0.069   &   0.067   &   0.044   &   0.062   &   7.728
 \\ \hline2009  &   0.279   &   0.203   &   0.123   &   0.077   &   0.079   &   0.072   &   0.064   &   0.051   &   0.051   &   4.421
 \\ \hline2010  &   0.282   &   0.197   &   0.131   &   0.074   &   0.085   &   0.056   &   0.079   &   0.038   &   0.056   &   9.723
 \\ \hline2011  &   0.290   &   0.195   &   0.126   &   0.079   &   0.077   &   0.067   &   0.069   &   0.038   &   0.059   &   5.761
 \\ \hline &\multicolumn{9}{|c|}{Lazio,  $N_c = $ 378 } &
 % \\ \hline Year/Digit &   1.000   &   2.000   &   3.000   &   4.000   &   5.000   &   6.000   &   7.000   &   8.000   &   9.000   &       &$ \chi^2 $\\ \hline
 \\ \hline2007  &   0.304   &   0.193   &   0.119   &   0.095   &   0.061   &   0.058   &   0.053   &   0.056   &   0.061   &   4.980
 \\ \hline2008  &   0.310   &   0.188   &   0.127   &   0.093   &   0.069   &   0.050   &   0.066   &   0.042   &   0.056   &   4.360
 \\ \hline2009  &   0.315   &   0.193   &   0.116   &   0.087   &   0.082   &   0.048   &   0.061   &   0.040   &   0.058   &   5.894
 \\ \hline2010  &   0.320   &   0.190   &   0.119   &   0.093   &   0.077   &   0.053   &   0.056   &   0.034   &   0.058   &   5.613
 \\ \hline2011  &   0.312   &   0.198   &   0.127   &   0.079   &   0.077   &   0.061   &   0.045   &   0.048   &   0.053   &   4.297
 \\ \hline
  \end{tabular}
\caption{Frequencies of the first digit in  reported AIT  data for 
IT regions with a high $N_c$ number, for various years;   it can be
visually compared to  the expected frequency according to BL given
in Table \ref {tableBL1}; the corresponding calculated  $\chi^2$ is
to be compared with the theoretical one (15.507)  for  a  number of
degree of freedom  ($\partial =8$)  at $0.05 \%$. The null
hypothesis is not verified for underlined cases. }
\label{TableNcityperregion1}
\end{center} \end{table}
  \begin{table} \begin{center}
\begin{tabular}[t]{|c|c|c|c|c|c|c|c|c|c|c|}
  \hline Digit  &   1   &   2   &   3   &   4   &   5   &   6   &   7   &   8   &   9   &$ \chi^2 $\\ \hline\hline
 Year  &\multicolumn{9}{|c|}{Sardegna,      $N_c = $ 337 } &
 % \\ \hline Year/Digit &   1.000   &   2.000   &   3.000   &   4.000   &   5.000   &   6.000   &   7.000   &   8.000   &   9.000   &       &$ \chi^2 $\\ \hline
 \\ \hline2007  &   0.310   &   0.167   &   0.093   &   0.053   &   0.050   &   0.119   &   0.064   &   0.109   &   0.034   &   \underline{56.001}
 \\ \hline2008  &   0.310   &   0.172   &   0.095   &   0.106   &   0.069   &   0.093   &   0.077   &   0.024   &   0.053   &   \underline{15.606}
 \\ \hline2009  &   0.316   &   0.175   &   0.101   &   0.095   &   0.085   &   0.069   &   0.093   &   0.032   &   0.034   &   13.908
 \\ \hline2010  &   0.318   &   0.175   &   0.095   &   0.090   &   0.093   &   0.061   &   0.095   &   0.034   &   0.037   &   \underline{16.059}
 \\ \hline2011  &   0.314   &   0.174   &   0.103   &   0.087   &   0.092   &   0.066   &   0.103   &   0.029   &   0.032   &   \underline{21.363}
  \\  \hline  &\multicolumn{9}{|c|}{Emilia-Romagna,     $N_c = $ 348 } &
 % \\ \hline Year/Digit &   1.000   &   2.000   &   3.000   &   4.000   &   5.000   &   6.000   &   7.000   &   8.000   &   9.000   &       &$ \chi^2 $\\ \hline
 \\ \hline2007  &   0.296   &   0.201   &   0.103   &   0.078   &   0.069   &   0.069   &   0.057   &   0.066   &   0.060   &   7.516
 \\ \hline2008  &   0.310   &   0.204   &   0.101   &   0.080   &   0.066   &   0.066   &   0.060   &   0.063   &   0.049   &   6.121
 \\ \hline2009  &   0.305   &   0.201   &   0.103   &   0.072   &   0.075   &   0.060   &   0.060   &   0.060   &   0.063   &   8.040
 \\ \hline2010  &   0.290   &   0.201   &   0.112   &   0.072   &   0.072   &   0.055   &   0.066   &   0.063   &   0.069   &   10.604
 \\ \hline2011  &   0.299   &   0.195   &   0.115   &   0.072   &   0.060   &   0.069   &   0.063   &   0.072   &   0.055   &   8.529
 \\ \hline &\multicolumn{9}{|c|}{Trentino-Alto Adige,      $N_c = $ 339 } &
 % \\ \hline Year/Digit &   1.000   &   2.000   &   3.000   &   4.000   &   5.000   &   6.000   &   7.000   &   8.000   &   9.000   &       &$ \chi^2 $\\ \hline
 \\ \hline2007  &   0.322   &   0.159   &   0.127   &   0.100   &   0.077   &   0.065   &   0.062   &   0.062   &   0.027   &   4.713
 \\ \hline2008  &   0.300   &   0.168   &   0.126   &   0.099   &   0.081   &   0.048   &   0.075   &   0.060   &   0.042   &   4.225
 \\ \hline2009  &   0.285   &   0.171   &   0.135   &   0.105   &   0.078   &   0.048   &   0.084   &   0.039   &   0.054   &   7.975
 \\ \hline2010  &   0.300   &   0.168   &   0.120   &   0.114   &   0.078   &   0.057   &   0.072   &   0.048   &   0.042   &   2.992
 \\ \hline2011  &   0.294   &   0.156   &   0.117   &   0.120   &   0.087   &   0.048   &   0.072   &   0.048   &   0.057   &   6.986
 \\ \hline &\multicolumn{9}{|c|}{Abruzzo,   $N_c = $ 305 } &
 % \\ \hline Year/Digit &   1.000   &   2.000   &   3.000   &   4.000   &   5.000   &   6.000   &   7.000   &   8.000   &   9.000   &       &$ \chi^2 $\\ \hline
 \\ \hline2007  &   0.292   &   0.187   &   0.105   &   0.085   &   0.089   &   0.056   &   0.075   &   0.052   &   0.059   &   5.381
 \\ \hline2008  &   0.295   &   0.170   &   0.111   &   0.089   &   0.098   &   0.059   &   0.062   &   0.056   &   0.059   &   3.852
 \\ \hline2009  &   0.289   &   0.164   &   0.111   &   0.102   &   0.079   &   0.062   &   0.066   &   0.056   &   0.072   &   6.090
 \\ \hline2010  &   0.298   &   0.154   &   0.121   &   0.092   &   0.085   &   0.069   &   0.066   &   0.052   &   0.062   &   3.252
 \\ \hline2011  &   0.318   &   0.144   &   0.115   &   0.095   &   0.079   &   0.062   &   0.079   &   0.059   &   0.049   &   5.111
 \\ \hline &\multicolumn{9}{|c|}{Toscana,   $N_c = $ 287 } &
 % \\ \hline Year/Digit &   1.000   &   2.000   &   3.000   &   4.000   &   5.000   &   6.000   &   7.000   &   8.000   &   9.000   &       &$ \chi^2 $\\ \hline
 \\ \hline2007  &   0.331   &   0.188   &   0.125   &   0.059   &   0.066   &   0.070   &   0.035   &   0.059   &   0.066   &   11.580
 \\ \hline2008  &   0.328   &   0.195   &   0.105   &   0.080   &   0.070   &   0.059   &   0.042   &   0.059   &   0.063   &   7.097
 \\ \hline2009  &   0.321   &   0.206   &   0.094   &   0.084   &   0.073   &   0.056   &   0.038   &   0.066   &   0.063   &   10.148
 \\ \hline2010  &   0.341   &   0.195   &   0.101   &   0.084   &   0.066   &   0.059   &   0.042   &   0.045   &   0.066   &   8.958
 \\ \hline2011  &   0.369   &   0.185   &   0.101   &   0.091   &   0.073   &   0.042   &   0.056   &   0.038   &   0.045   &   9.787
 \\ \hline &\multicolumn{9}{|c|}{Puglia,    $N_c = $ 258 } &
 % \\ \hline Year/Digit &   1.000   &   2.000   &   3.000   &   4.000   &   5.000   &   6.000   &   7.000   &   8.000   &   9.000   &       &$ \chi^2 $\\ \hline
 \\ \hline2007  &   0.306   &   0.194   &   0.116   &   0.089   &   0.089   &   0.043   &   0.054   &   0.047   &   0.062   &   5.060
 \\ \hline2008  &   0.295   &   0.213   &   0.101   &   0.101   &   0.074   &   0.078   &   0.039   &   0.047   &   0.054   &   5.989
 \\ \hline2009  &   0.291   &   0.217   &   0.101   &   0.097   &   0.074   &   0.078   &   0.035   &   0.062   &   0.047   &   7.260
 \\ \hline2010  &   0.302   &   0.209   &   0.105   &   0.101   &   0.078   &   0.074   &   0.031   &   0.062   &   0.039   &   6.800
 \\ \hline2011  &   0.310   &   0.202   &   0.112   &   0.101   &   0.070   &   0.074   &   0.035   &   0.054   &   0.043   &   4.325
 \\ \hline
  \end{tabular}
\caption{Frequencies of the first digit in  reported AIT  data for
IT regions with a medium range $N_c$ number, for various years;
it can be visually compared to  the expected frequency according to
BL given in Table \ref {tableBL1}; the corresponding calculated
$\chi^2$ is to be compared with the theoretical one (15.507)  for  a
number of degree of freedom  ($\partial =8$)  at $0.05 \%$. The null
hypothesis is  always verified in these cases. }
\label{TableNcityperregion2}
\end{center} \end{table}
  \begin{table} \begin{center}
\vspace{-50pt}
  \begin{tabular}[t]{|c|c|c|c|c|c|c|c|c|c|c|}
  \hline Digit  &   1   &   2   &   3   &   4   &   5   &   6   &   7   &   8   &   9   &$ \chi^2 $\\ \hline\hline
   Year&\multicolumn{9}{|c|}{Marche,    $N_c = $ 239 } &
 % \\ \hline Year/Digit &   1.000   &   2.000   &   3.000   &   4.000   &   5.000   &   6.000   &   7.000   &   8.000   &   9.000   &       &$ \chi^2 $\\ \hline
 \\ \hline2007  &   0.268   &   0.172   &   0.092   &   0.121   &   0.084   &   0.063   &   0.054   &   0.071   &   0.075   &   11.051
 \\ \hline2008  &   0.285   &   0.172   &   0.092   &   0.113   &   0.084   &   0.071   &   0.059   &   0.054   &   0.071   &   6.486
 \\ \hline2009  &   0.280   &   0.180   &   0.088   &   0.117   &   0.088   &   0.071   &   0.067   &   0.042   &   0.067   &   7.370
 \\ \hline2010  &   0.285   &   0.180   &   0.096   &   0.096   &   0.109   &   0.054   &   0.071   &   0.038   &   0.071   &   9.946
 \\ \hline2011  &   0.293   &   0.197   &   0.088   &   0.096   &   0.105   &   0.042   &   0.071   &   0.054   &   0.054   &   8.608
 \\ \hline &\multicolumn{9}{|c|}{Liguria,   $N_c = $ 235 } &
 % \\ \hline Year/Digit &   1.000   &   2.000   &   3.000   &   4.000   &   5.000   &   6.000   &   7.000   &   8.000   &   9.000   &       &$ \chi^2 $\\ \hline
 \\ \hline2007  &   0.272   &   0.209   &   0.085   &   0.089   &   0.077   &   0.098   &   0.043   &   0.047   &   0.081   &   \underline{15.92}0
 \\ \hline2008  &   0.268   &   0.213   &   0.089   &   0.081   &   0.055   &   0.111   &   0.055   &   0.034   &   0.094   &   \underline{27.173}
 \\ \hline2009  &   0.294   &   0.209   &   0.077   &   0.098   &   0.060   &   0.094   &   0.060   &   0.034   &   0.077   &   \underline{15.719}
 \\ \hline2010  &   0.289   &   0.196   &   0.094   &   0.102   &   0.047   &   0.094   &   0.072   &   0.030   &   0.077   &   \underline{15.954}
 \\ \hline2011  &   0.306   &   0.191   &   0.089   &   0.115   &   0.047   &   0.089   &   0.060   &   0.043   &   0.060   &   9.707
 \\ \hline &\multicolumn{9}{|c|}{Friuli-Venezia Giulia,      $N_c = $ 219 for 2007  ;  $N_c = $ 218  for 2008-2011 } &
 % \\ \hline Year/Digit &   1.000   &   2.000   &   3.000   &   4.000   &   5.000   &   6.000   &   7.000   &   8.000   &   9.000   &       &$ \chi^2 $\\ \hline
 \\  \hline2007     &   0.279   &   0.210   &   0.142   &   0.078   &   0.055   &   0.059   &   0.068   &   0.064   &   0.046   &   6.075
 \\ \hline2008  &   0.289   &   0.220   &   0.138   &   0.078   &   0.046   &   0.073   &   0.055   &   0.064   &   0.037   &   7.940
 \\ \hline2009  &   0.280   &   0.220   &   0.142   &   0.078   &   0.041   &   0.069   &   0.064   &   0.064   &   0.041   &   8.993
 \\ \hline2010  &   0.303   &   0.206   &   0.142   &   0.073   &   0.050   &   0.073   &   0.060   &   0.060   &   0.032   &   6.516
 \\ \hline2011  &   0.284   &   0.211   &   0.151   &   0.069   &   0.055   &   0.073   &   0.041   &   0.073   &   0.041   &   9.698
 \\ \hline &\multicolumn{9}{|c|}{Molise,    $N_c = $ 136 } &
 % \\ \hline Year/Digit &   1.000   &   2.000   &   3.000   &   4.000   &   5.000   &   6.000   &   7.000   &   8.000   &   9.000   &       &$ \chi^2 $\\ \hline
 \\ \hline2007  &   0.243   &   0.125   &   0.191   &   0.118   &   0.074   &   0.074   &   0.074   &   0.051   &   0.051   &   9.741
 \\ \hline2008  &   0.221   &   0.110   &   0.213   &   0.118   &   0.059   &   0.103   &   0.044   &   0.074   &   0.059   &   \underline{20.990}
 \\ \hline2009  &   0.250   &   0.110   &   0.184   &   0.110   &   0.081   &   0.103   &   0.051   &   0.066   &   0.044   &   11.890
 \\ \hline2010  &   0.228   &   0.132   &   0.169   &   0.118   &   0.096   &   0.081   &   0.066   &   0.037   &   0.074   &   10.475
 \\ \hline2011  &   0.235   &   0.140   &   0.169   &   0.096   &   0.103   &   0.081   &   0.081   &   0.029   &   0.066   &   10.190
 \\ \hline &\multicolumn{9}{|c|}{Basilicata,    $N_c = $ 131 } &
 % \\ \hline Year/Digit &   1.000   &   2.000   &   3.000   &   4.000   &   5.000   &   6.000   &   7.000   &   8.000   &   9.000   &       &$ \chi^2 $\\ \hline
 \\ \hline2007  &   0.290   &   0.160   &   0.130   &   0.061   &   0.122   &   0.076   &   0.061   &   0.023   &   0.076   &   9.965
 \\ \hline2008  &   0.282   &   0.168   &   0.107   &   0.107   &   0.084   &   0.092   &   0.031   &   0.061   &   0.069   &   5.365
 \\ \hline2009  &   0.290   &   0.160   &   0.122   &   0.099   &   0.061   &   0.099   &   0.053   &   0.053   &   0.061   &   3.567
 \\ \hline2010  &   0.290   &   0.168   &   0.115   &   0.092   &   0.076   &   0.122   &   0.023   &   0.053   &   0.061   &   9.692
 \\ \hline2011  &   0.305   &   0.160   &   0.099   &   0.115   &   0.053   &   0.122   &   0.046   &   0.046   &   0.053   &   8.938
 \\ \hline &\multicolumn{9}{|c|}{Umbria,    $N_c = $ 92  } &
 % \\ \hline Year/Digit &   1.000   &   2.000   &   3.000   &   4.000   &   5.000   &   6.000   &   7.000   &   8.000   &   9.000   &       &$ \chi^2 $\\ \hline
 \\ \hline2007  &   0.380   &   0.207   &   0.130   &   0.065   &   0.109   &   0.054   &   0.011   &   0.022   &   0.022   &   10.855
 \\ \hline2008  &   0.370   &   0.207   &   0.130   &   0.065   &   0.120   &   0.043   &   0.022   &   0.022   &   0.022   &   10.348
 \\ \hline2009  &   0.370   &   0.207   &   0.130   &   0.076   &   0.120   &   0.043   &   0.022   &   0.022   &   0.011   &   11.093
 \\ \hline2010  &   0.370   &   0.217   &   0.120   &   0.076   &   0.109   &   0.065   &   0.011   &   0.000   &   0.033   &   12.352
 \\ \hline2011  &   0.380   &   0.217   &   0.109   &   0.087   &   0.109   &   0.054   &   0.011   &   0.011   &   0.022   &   11.938
 \\ \hline &\multicolumn{9}{|c|}{Valle d'Aosta,     $N_c = $ 74 } &
 % \\ \hline Year/Digit &   1.000   &   2.000   &   3.000   &   4.000   &   5.000   &   6.000   &   7.000   &   8.000   &   9.000   &       &$ \chi^2 $\\ \hline
 \\ \hline2007  &   0.270   &   0.230   &   0.095   &   0.081   &   0.122   &   0.054   &   0.081   &   0.041   &   0.027   &   5.456
 \\ \hline2008  &   0.270   &   0.243   &   0.108   &   0.068   &   0.108   &   0.054   &   0.081   &   0.014   &   0.054   &   6.760
 \\ \hline2009  &   0.284   &   0.203   &   0.122   &   0.054   &   0.149   &   0.054   &   0.068   &   0.027   &   0.041   &   7.477
 \\ \hline2010  &   0.284   &   0.216   &   0.108   &   0.027   &   0.162   &   0.068   &   0.054   &   0.041   &   0.041   &   11.309
 \\ \hline2011  &   0.284   &   0.176   &   0.135   &   0.041   &   0.135   &   0.081   &   0.054   &   0.027   &   0.068   &   7.339
  \\ \hline
\end{tabular}
\caption{Frequencies of the first digit in  reported AIT  data for
IT regions with a low   $N_c$ number, for various years;   it can be
visually compared to  the expected frequency according to BL given
in Table \ref {tableBL1}; the corresponding calculated  $\chi^2$ is
to be compared with the theoretical one (15.507)  for  a  number of
degree of freedom  ($\partial =8$)  at $0.05 \%$. The null
hypothesis is not verified for underlined cases. }
\label{TableNcityperregion3}
\end{center} \end{table}

\end{document}